\documentclass[11pt]{article}

\usepackage[utf8]{inputenc}
\usepackage[T1]{fontenc}
\usepackage{mathptmx}
\usepackage[a4paper,margin=1in]{geometry}
\usepackage{setspace}
\usepackage{microtype}
\usepackage{amsmath,amssymb,amsthm,mathtools}
\usepackage{booktabs,array,multirow,longtable}
\usepackage[round,authoryear]{natbib}
\usepackage{xcolor}
\usepackage{enumitem}
\usepackage{fancyhdr}
\usepackage{titlesec}
\usepackage{hyperref}

\definecolor{brandburgundy}{RGB}{128,0,32}
\hypersetup{
  colorlinks=true,
  linkcolor=brandburgundy,
  citecolor=brandburgundy,
  urlcolor=brandburgundy,
  pdftitle={The Replicator-Optimization Mechanism: A Scale-Relative Formalism for Persistence-Conditioned Dynamics with a Conditional Consent-Friction Instantiation},
  pdfauthor={Murad Farzulla}
}

\newcommand{\paperwp}{DAI-2503}
\newcommand{\paperdate}{12 September 2026}
\newcommand{\paperurl}{https://systems.ac/4/DAI-2503}
\newcommand{\ROM}{ROM}

\fancypagestyle{firstpage}{\fancyhf{}\fancyfoot[C]{\small\thepage}}

\titleformat{\section}{\normalfont\large\bfseries\color{brandburgundy}}{\thesection}{0.5em}{}
\titleformat{\subsection}{\normalfont\normalsize\bfseries\color{brandburgundy}}{\thesubsection}{0.5em}{}
\titleformat{\subsubsection}{\normalfont\small\bfseries\color{brandburgundy}}{\thesubsubsection}{0.5em}{}
\titlespacing*{\section}{0pt}{2ex plus 0.8ex minus 0.2ex}{1ex plus 0.3ex}
\titlespacing*{\subsection}{0pt}{1.5ex plus 0.5ex minus 0.2ex}{0.8ex plus 0.2ex}
\titlespacing*{\subsubsection}{0pt}{1.2ex plus 0.4ex minus 0.2ex}{0.6ex plus 0.2ex}

\theoremstyle{definition}
\newtheorem{definition}{Definition}[section]

\theoremstyle{plain}
\newtheorem{proposition}{Proposition}[section]

\newtheorem{theorem}{Theorem}[section]
\newtheorem{corollary}{Corollary}[section]

\theoremstyle{remark}
\newtheorem{remark}{Remark}[section]

\begin{document}
\setstretch{1.2}
\thispagestyle{firstpage}

\begin{center}
{\small\textsc{\href{https://dissensus.ai}{Dissensus} Working Paper Series}}\\[0.2em]
{\small \href{\paperurl}{\paperwp}}\\[1.5em]
{\LARGE\bfseries The Replicator-Optimization Mechanism:}\\[0.3em]
{\large\bfseries A Scale-Relative Formalism for Persistence-Conditioned Dynamics}\\[0.6em]
{\normalsize\itshape with a Conditional Consent-Friction Instantiation}\\[1.5em]
{\large Murad Farzulla}\textsuperscript{1,2,*}\\[0.8em]
{\small \textsuperscript{1}\href{https://dissensus.ai}{Dissensus}, London, UK \quad
\textsuperscript{2}King's College London, London, UK}\\[0.5em]
{\footnotesize \textsuperscript{*}Correspondence: \href{mailto:murad@dissensus.ai}{murad@dissensus.ai}
\quad ORCID: \href{https://orcid.org/0009-0002-7164-8704}{0009-0002-7164-8704}}\\[0.3em]
{\footnotesize \paperdate}
\end{center}

\begin{abstract}
\noindent Persistence models often conflate propagation, survival, and cross-scale loss. The Replicator-Optimization Mechanism (ROM) is a replicator--mutator template separating baseline weight, bounded survival, and a transfer kernel at a declared scale. Its equation conserves mass but guarantees neither invariance, convergence, a potential, nor a preferred scale. For finite static density-independent continuous time, an irreducible weighted kernel yields a unique positive Perron--Frobenius composition; discrete-time power convergence needs primitivity. The componentwise ranking proved here is guaranteed under exact uniform-residual transfer. Strong lumpability gives universal first-order transfer closure, and blockwise effective fitness gives an exact quotient.

An institutional instantiation uses normalized stakes, signed preference--decision alignment, information loss, and descriptive effective voice. It specifies conditional survival, not legitimacy or normative authority. A companion mixed-motive MARL battery reports exploratory evidence against the implemented proxy ratio in its environment: a positive signed target-coordinate-correlation effect survives held-out evaluation under shared-state contention, while a reduced feasible-centred frozen-policy crossing reverses the predicted correlation--noise interaction. The treatment varies ideal-point correlation inside a fixed reward family, not objective- or reward-function alignment. Lean checks mapped algebraic identities and scalar monotonicities, not the stationary theorem, empirical mapping, or normative bridge. ROM is an assumptions ledger and model-construction discipline, not a cross-substrate law.

\vspace{0.8em}
\noindent\textbf{Keywords:} persistence, replicator--mutator dynamics, scale, lumpability, effective fitness, institutional modeling, formal verification
\end{abstract}

\vspace{0.4em}
\noindent\textbf{Ancillary files:} The standalone technical supplement is included in the arXiv source package as \texttt{anc/technical-supplement\_ROM.pdf}; the exact cited Lean modules remain under \texttt{anc/lean/}.

\newpage

\section{Introduction}
\label{sec:introduction}

Persistence is not one mechanism. A configuration may remain common because it reproduces quickly, because it survives conditional on appearing, because a transfer process continually recreates it, or because the state description hides the relevant replacements. Treating all four cases as ``fitness'' after observing survival makes the explanation circular. A useful persistence model instead declares its units, transition law, and effective-fitness components before comparing trajectories.

The Replicator-Optimization Mechanism (\ROM{}) is a notation and workflow for doing that. At a declared scale $S$, a modeler specifies a finite type space, a baseline rate $w_S$, a bounded survival factor $\rho_S$, and a row-stochastic transfer kernel $M_S$. The resulting equation is a standard selection-with-transfer form. The name ``optimization'' refers only to differential amplification under stipulated effective-fitness rates. It does not assert utility maximization, informational optimality, global convergence, or that realized survivors were normatively best.

This modest description matters because the relevant mathematical traditions overlap without collapsing into one another. Replicator and Price-equation methods separate selection from transmission in population models \citep{taylor1978evolutionary,price1970selection,page2002unifying}. Evolutionary economics and cultural evolution adapt those operators to routines and socially transmitted traits \citep{nelsonwinter1982,elmouden2014cultural}, in institutional settings whose change dynamics have their own tradition \citep{north1990institutions}. Coarse-graining and projection theory ask when an aggregate process closes and when hidden state or memory remains \citep{mori1965transport,zwanzig1961memory,kemeny1976finite}. ROM borrows this common discipline---state the level, law, and closure assumptions---without claiming that social persistence is thermodynamic dissipation or that every substrate shares a causal mechanism.

The paper makes four contributions. First, it states the finite-type ROM equation with a clear boundary between identities and dynamical theorems. Second, it transfers to ROM the restricted stationary result that its companion normative paper previously carried: a static density-independent continuous-time system with an irreducible weighted kernel converges to its unique positive Perron--Frobenius composition, while typewise ranking by effective fitness is proved here under an exact uniform-residual kernel. Third, it applies strong lumpability to state the exact cross-scale closure condition and refuses any universal direct-versus-stepwise approximation ordering outside that case. Fourth, it gives a conditional institutional instantiation and records what its empirical control battery supports and contradicts.

The division of labor across the companion papers is deliberate. The Axiom of Consent (AOC) supplies a proposed normative authorization rule and the philosophical argument needed before descriptive effective voice can be called legitimacy \citep{farzulla2025aoc}. ROM owns the persistence dynamics and their restrictions. The MARL control battery owns the experimental construct audit of the proposed friction index \citep{farzulla2026marl}. None inherits the others' conclusions automatically: a normative premise is not a survival theorem, a survival equation is not a measurement model, and a result in a stylized resource game is not evidence of institutional convergence.

Section~\ref{sec:scope} fixes the objects and scope. Section~\ref{sec:dynamics} states the dynamics, stationary theorem, kernel correspondences, and scale criterion. Section~\ref{sec:institutional} gives the institutional instantiation. Section~\ref{sec:operational} separates measurement, identification, and current evidence. Section~\ref{sec:example} works one small example, and Section~\ref{sec:discussion} closes the formal and normative boundaries. Technical motivations, proofs, network diagnostics, and the Lean map appear in the appendices. The ROM technical supplement retains the full closure, network, ownership, and Lean materials; the AOC technical pointers \texttt{TS-FORM} and \texttt{TS-MEASURE} retain the longer friction-index and measurement motivations.

\section{Scope and modeling commitments}
\label{sec:scope}

\begin{definition}[Declared scale and type space]
A scale $S$ is a modeling resolution with a specified observation interval and a finite set $T_S$ of distinguishable types. A type is the unit whose relative mass the equation tracks. It may represent a strategy, routine, organizational form, or other application-specific state; calling it an agent is not required.
\end{definition}

Scale is therefore a measurement choice, not an ontological claim. The same physical or social system can support multiple descriptions, and their state spaces need not be nested in a unique way. A useful scale is one whose retained variables support the prediction or explanation being attempted. Section~\ref{sec:coarse-graining} supplies an exact closure test; outside that test, usefulness is assessed against a declared loss and horizon rather than inferred from granularity.

The declaration has to identify an observation map as well as a label. Let $X_t$ denote whatever lower-level configuration is available and let $r_S(X_t)\in T_S$ be the rule by which it is classified. The population coordinate
\[
p_t(\tau)=\Pr\bigl(r_S(X_t)=\tau\bigr)
\]
may be an ensemble frequency, a large-population share, or an empirical distribution over repeated units. These interpretations cannot be exchanged silently. A distribution across institutions at one date, for example, is not the time trajectory of a single institution. Nor is a transition between labels necessarily reproduction: it may be replacement, imitation, reform, or reclassification. The application must state which event moves mass and which events leave the type label unchanged.

\begin{definition}[Autonomous, nonautonomous, and coupled specifications]
A ROM specification is \emph{autonomous} when its vector field is a fixed function of $p$; it is \emph{nonautonomous} when a declared input such as $G_t$ or $M_t$ varies with time; and it is \emph{coupled} when those inputs obey additional equations jointly with $p$. Writing a time subscript on an input specifies neither its path nor a law for its evolution.
\end{definition}

This distinction fixes what counts as an equilibrium. In an autonomous model, $p^*$ is stationary when the specified vector field vanishes at $p^*$. In a nonautonomous model, an instantaneous zero of the vector field can move with the inputs and is not a fixed point of the resulting trajectory. A coupled model may have an equilibrium in the enlarged state $(p,G,\theta)$ even when the marginal $p$ equation alone does not close. ROM contains only the $p$ equation until the application supplies the other laws. In particular, it does not imply adaptation of $G$, updating of the scores, or co-evolution of institutions and affected parties.

ROM applies only after five commitments are made:
\begin{enumerate}[noitemsep]
    \item \textbf{State.} The type space and population distribution are observable or estimable at the selected scale.
    \item \textbf{Effective fitness.} Baseline weight and conditional survival combine into a rate with declared units and covariates.
    \item \textbf{Transfer.} A nonnegative row-stochastic kernel describes mutation, learning, imitation, reform, or another named transition process.
    \item \textbf{Environment.} Network and exogenous variables are stated as fixed, time-varying inputs, or explicit dynamical states. Merely writing $G_{S,t}$ does not supply a co-evolution law for $G$.
    \item \textbf{Approximation regime.} A deterministic frequency equation requires a population or sampling argument appropriate to the dependence and tail structure. Large size alone is not a proof of concentration.
\end{enumerate}

The domains of the objects are part of those commitments. Throughout ROM, $p$ lies in the finite probability simplex; $M$ is a nonnegative row-stochastic matrix; $w$ is a nonnegative rate; and the survival factor obeys $0\leq\rho\leq1$. The bound makes $\rho$ an attenuation of baseline capacity rather than a second unrestricted rate scale. The general stationary theorem below uses the strictly positive interior $0<\rho\leq1$ and positive static $w$. The institutional choice $\rho=L/(1+F)$ lies in the general domain $0\leq\rho\leq1$ because $0\leq L\leq1$ and $F\geq0$. It meets the stationary theorem's strictly positive requirement only when $L>0$ for every type; a type with zero effective voice contributes no fitness-weighted outgoing mass and can break the irreducibility of $Q$ (Section~\ref{sec:dynamics} returns to the zero case before Theorem~\ref{thm:rom-stationary}).

The distinction between type and bearer is equally important. ROM tracks mass assigned to types, not the moral status, welfare, or intention of the entities bearing them. An institutional type can outlast every current officeholder; an individual strategy can disappear while the individual remains; a policy can be copied without any new organization being born. These are different event ontologies represented by the same mathematical slot. The common notation is useful only if the application preserves the event interpretation when it defines $w$ and $M$.

The finite type space is a substantive restriction. Continuous traits, unbounded state spaces, and measure-valued populations require additional topology, regularity, and existence results. Likewise, ``entropy'' outside a thermodynamic model is not a free-standing force. In the institutional application below, $\varepsilon$ is a declared information-loss coordinate. It has empirical meaning only through its estimator and data-generating process.

The deterministic frequency equation is also an approximation whenever the underlying units are finite and stochastic. Independence is not required in every limit theorem, but some control of dependence and tails is. Common shocks, network clusters, path-dependent replacement, or a few dominant units can prevent concentration even in a numerically large population. A responsible application should therefore name the sampling unit, effective sample size, and stochastic comparator. If demographic noise is substantively large, a master equation, finite-population Markov process, or stochastic differential equation may be the relevant object; ROM's ordinary differential equation does not make that noise disappear.

These commitments keep three levels of claim separate. A \emph{definition} fixes notation. An \emph{identity} follows algebraically from that notation. A \emph{dynamical result} requires conditions on the vector field or kernel. Cross-domain similarity can motivate a specification, but approximate validity on a new substrate requires its own error or predictive comparison.

\section{ROM dynamics}
\label{sec:dynamics}

\subsection{The finite-type equation}

For $\tau\in T_S$, let $p_t(\tau)$ denote the type's population share. Define effective fitness
\begin{equation}
\phi_S(\tau,G,p,t):=w_S(\tau,t)\rho_S(\tau,G,p,t),
\qquad
\bar\phi_t:=\sum_{\tau}p_t(\tau)\phi_S(\tau,G_{S,t},p_t,t).
\label{eq:effective-fitness}
\end{equation}
The ROM template is the continuous-time weighted replicator--mutator equation
\begin{equation}
\frac{dp_t(\tau)}{dt}
=\sum_{\tau'\in T_S}p_t(\tau')\phi_S(\tau',G_{S,t},p_t,t)
M_S(\tau'\to\tau,t)
-p_t(\tau)\bar\phi_t.
\label{eq:rom-main}
\end{equation}
Here $M_S$ is nonnegative and row-stochastic. In continuous time, $w_S$ and $\phi_S$ have units of inverse time, whereas $\rho_S$ and $M_S$ are dimensionless. The split $\phi=w\rho$ is conceptually useful but not identified from the population path without further information (Section~\ref{sec:identifiability}).

The word survival names a model component, not automatically an observed event probability. In continuous time, $\rho$ attenuates the baseline rate $w$; converting a measured one-period survival probability or hazard into that factor requires a time convention and link function. A value of $\rho=.8$ does not mean that 80 percent of units survive the paper's observation window unless the application has defined that mapping. Conversely, the bound $\rho\leq1$ does not bound effective fitness unless $w$ is bounded. Reporting the score without its baseline rate and time unit is therefore insufficient to reproduce the vector field.

Zeros are allowed in the general ROM domain but excluded from Theorem~\ref{thm:rom-stationary}. If $\rho(\tau)=0$, the type contributes no fitness-weighted outgoing mass through its own row of $Q$, although it may still receive inflow from other types. Such a zero can change the communicating structure of the weighted operator and invalidate the positive-eigenvector argument. It is not necessarily an absorbing state: absorption is a property of transition and contribution together. Empirical censoring at zero should therefore be distinguished from a structurally estimated zero.

In row-vector notation, with $Q(p,t)=\operatorname{diag}(\phi(p,t))M(t)$ and $\mathbf 1$ the all-ones column vector, the equation is
\begin{equation}
\dot p=pQ-p\bigl(pQ\mathbf1\bigr).
\label{eq:rom-vector}
\end{equation}
This form separates a positive inflow operator from the normalization that keeps total mass fixed. It also shows why $M$ is not an additional population distribution: its rows are conditional destination probabilities given a source type, while $p$ supplies the source mass. The diagonal factor is attached to the source before transfer. A destination-weighted convention would be a different model.

Summing Equation~\eqref{eq:rom-main} over destinations gives
\[
\sum_\tau\dot p_t(\tau)
=\sum_{\tau'}p_t(\tau')\phi_S(\tau')\sum_\tau M_S(\tau'\to\tau)
-\bar\phi_t\sum_\tau p_t(\tau)=0
\]
when $p_t$ is normalized. This is total-mass conservation. It is not, by itself, an existence theorem or a proof that a solution beginning in the simplex remains componentwise nonnegative. The Lean formalization checks this row-sum identity, not the omitted differential-equation argument.

For a finite locally Lipschitz specification, nonnegative $Q$, and a solution that exists, the boundary points inward: if $p_i=0$, Equation~\eqref{eq:rom-vector} gives $\dot p_i=(pQ)_i\geq0$. Together with the zero-sum identity, standard invariance results can then keep the trajectory in the simplex. Those analytic premises are short to state but are not formalized in the accompanying Lean development, so the paper does not package their conclusion as machine-checked. Unbounded or discontinuous score functions can also destroy the global-existence step even though the row-sum algebra remains true.

When $G$, $w$, $\rho$, and $M$ are fixed in time and $\rho$ is density-independent, Equation~\eqref{eq:rom-main} is an autonomous normalized positive linear flow. The next theorem states what follows in that restricted case. In a time-varying or frequency-dependent system, it does not apply.

\subsection{Restricted stationary result}
\label{sec:static-theorem}

\begin{theorem}[Finite static continuous-time ROM convergence]
\label{thm:rom-stationary}
Let $T_S$ be finite. Suppose $w_S(\tau)>0$ and $0<\rho_S(\tau)\leq1$ are time-independent and independent of $p$, and let $M_S$ be a time-independent row-stochastic kernel. Define the nonnegative weighted matrix
\[
Q_{\tau'\tau}:=w_S(\tau')\rho_S(\tau')M_S(\tau'\to\tau).
\]
If $Q$ is irreducible---equivalently here, if $M_S$ is irreducible, because all $w_S\rho_S$ are positive---then every initial distribution $p_0$ converges under Equation~\eqref{eq:rom-main} to the unique normalized positive left Perron--Frobenius eigenvector $p^*$ of $Q$:
\begin{equation}
p^*Q=\lambda_*p^*,
\qquad \sum_\tau p^*(\tau)=1,
\label{eq:rom-stationary}
\end{equation}
where $\lambda_*>0$ is the Perron root and the unique eigenvalue with maximal real part.

If, in addition,
\begin{equation}
M_S=(1-\mu)I+\mu U,
\qquad 0<\mu<1,
\qquad U_{ij}=\frac{1}{|T_S|},
\label{eq:uniform-residual}
\end{equation}
then $p^*(\tau)$ is strictly increasing in the type's effective fitness $\phi(\tau)=w_S(\tau)\rho_S(\tau)$. This componentwise ranking need not hold for a general irreducible kernel.
\end{theorem}

\begin{proof}
Under the hypotheses, $Q$ is constant. Let the unnormalized row vector $y(t)$ solve $\dot y=yQ$ with $y(0)=p_0$, and write $z(t)=y(t)\mathbf 1$ and $p(t)=y(t)/z(t)$. Because $Q$ is nonnegative, $e^{tQ}$ is nonnegative and $z(t)>0$ for every $t\geq0$ when $p_0$ is a nonzero nonnegative vector. Direct differentiation gives
\[
\dot p=pQ-p\bigl(pQ\mathbf 1\bigr),
\]
which is Equation~\eqref{eq:rom-main} because $Q\mathbf 1$ is the vector of effective-fitness rates. Irreducibility and Perron--Frobenius theory give a simple Perron root $\lambda_*>0$ and positive left and right Perron eigenvectors, each unique up to scaling. For any other eigenvalue $\lambda$, either $|\lambda|<\lambda_*$ or it is a nontrivial peripheral eigenvalue, and in either case $\operatorname{Re}\lambda<\lambda_*$. If $r>0$ is a right Perron vector, the coefficient of the Perron spectral projection is positive because $p_0r>0$. No diagonalizability assumption is needed: in a Jordan decomposition, every non-Perron generalized-eigenspace term is a finite sum of factors $t^k e^{\lambda t}$, and $t^k e^{\operatorname{Re}(\lambda)t}=o(e^{\lambda_*t})$ because $\operatorname{Re}\lambda<\lambda_*$. The Perron term therefore dominates $p_0e^{tQ}$ asymptotically. Normalization by $z(t)$ yields $p(t)\to p^*$, and the limiting normalization term is $\lambda_*=p^*Q\mathbf 1$.

For Equation~\eqref{eq:uniform-residual}, stationarity in component $\tau$ gives
\[
p^*(\tau)\lambda_*
=(1-\mu)\phi(\tau)p^*(\tau)
+\frac{\mu}{|T_S|}\sum_{\tau'}\phi(\tau')p^*(\tau').
\]
Let $c=|T_S|^{-1}\sum_{\tau'}\phi(\tau')p^*(\tau')>0$. Then
\begin{equation}
p^*(\tau)=\frac{\mu c}{\lambda_*-(1-\mu)\phi(\tau)}.
\label{eq:uniform-ranking}
\end{equation}
The positive rank-one perturbation makes $\lambda_*>(1-\mu)\phi(\tau)$ for every $\tau$, so the denominator is positive and decreases strictly as $\phi(\tau)$ rises. For a general kernel, stationary mass depends on fitness-weighted inflow from other types; a type with larger own fitness need not receive larger mass. This is why the ranking clause proved here is stated under Equation~\eqref{eq:uniform-residual}; the paper does not claim that specification is necessary.
\end{proof}

The proof gives more than existence of a stationary vector but less than a universal evolutionary ranking. It gives convergence of the \emph{normalized composition}; the unnormalized mass $y(t)$ generally grows at the Perron rate. At the limit, the population mean effective fitness equals $\lambda_*$ because $p^*Q\mathbf1=\lambda_*p^*\mathbf1$. This equality is a property of the fixed-matrix normalization. It is not a welfare statement and does not identify which factor in $w\rho$ caused the dominant eigenvalue.

The eigenvector is also not the ordinary stationary distribution of $M$ unless the weights make those objects coincide. The transfer kernel satisfies $M\mathbf1=\mathbf1$, but the weighted matrix satisfies $Q\mathbf1=\phi$ and is generally not stochastic. Its Perron eigenvalue need not equal one. Calling $p^*$ stationary means that reproduction, survival, transfer, and normalization balance in Equation~\eqref{eq:rom-main}; it does not mean that a single unit following $M$ has distribution $p^*$. This difference matters when an empirical transition matrix is estimated separately from population abundance.

The theorem is continuous-time even though its closure argument uses a matrix exponential. The auxiliary flow $y(t)=p_0e^{tQ}$ is not the discrete chain with transition matrix $Q$, because $Q$ need not be row-stochastic. Normalizing the positive linear flow generates the nonlinear mean-fitness subtraction in ROM. Irreducibility suffices here because nontrivial peripheral eigenvalues have smaller real part and are damped relative to the Perron exponential. For the corresponding normalized discrete-generation iteration $p_{k+1}=p_kQ/(p_kQ\mathbf1)$, irreducibility alone can permit periodic cycling; primitivity---irreducibility plus aperiodicity for the associated finite graph---is the standard stronger condition that makes normalized powers converge.

The positivity and irreducibility assumptions do real work. Reducibility can preserve dependence on the initially occupied communicating class. Zeros in effective fitness can remove paths from the weighted matrix even when the unweighted kernel is connected. A reducible constant $Q$ requires a class-specific spectral decomposition; neither a unique positive composition nor convergence from every nonzero nonnegative start follows in general.

The limit of any general ranking claim can be seen without constructing a complicated selection game. Take a positive two-state kernel whose identical rows are $(.01,.99)$. Its normalized stationary composition is $(.01,.99)$ for any positive effective-fitness vector, because every transfer redraws the destination from that row. Type 1 may have higher own effective fitness yet much lower stationary mass. The matrix is primitive; what fails is the uniform-residual structure in Equation~\eqref{eq:uniform-residual}. Thus Perron convergence does not imply componentwise ranking, and a fitted association between stationary abundance and $w\rho$ cannot be assumed without specifying transfer.

The residual parameter also marks two informative boundaries. As $\mu\downarrow0$, transfer approaches identity and the uniform mixing that makes $Q$ primitive disappears at the endpoint; if effective-fitness values differ, the normalized flow can concentrate on the maximal types and ties can preserve initial-condition dependence. As $\mu\uparrow1$, every source draws uniformly from destinations. At the endpoint the stationary composition is uniform regardless of typewise effective fitness, so strict ranking disappears even though the weighted matrix remains positive. The open interval in Theorem~\ref{thm:rom-stationary} is therefore not decorative: it retains both a type's own contribution and positive residual inflow.

Two invariances help interpret comparisons within the theorem. Multiplying every $w(\tau)$ by the same constant multiplies $Q$ and $\lambda_*$ by that constant but leaves the normalized eigenvector $p^*$ unchanged; it changes clock speed, not stationary composition. By contrast, multiplying one type's rate changes a row of $Q$, so both the Perron rate and stationary composition can change through the entire transfer network. Likewise, moving a common factor between $w$ and $\rho$ leaves $Q$ unchanged, whereas truncating $\rho$ at its bounds generally does not. These facts are algebraic and do not identify a causal channel.

For a differentiable family of irreducible nonnegative matrices $Q(\theta)$, the Perron eigenvalue is simple and its local sensitivity can be written
\begin{equation}
\frac{d\lambda_*}{d\theta}
=\frac{\ell^\top Q'(\theta)r}{\ell^\top r},
\label{eq:perron-sensitivity}
\end{equation}
where $\ell$ and $r$ are positive left and right Perron eigenvectors. The formula concerns the aggregate growth rate of the unnormalized flow. It does not give a componentwise abundance ordering, and the derivative of the normalized left eigenvector requires the rest of the spectrum. A claim that one score change raises every corresponding stationary mass therefore needs more structure, such as the uniform-residual calculation, rather than only positivity of Equation~\eqref{eq:perron-sensitivity}.

The theorem is not a generic convergence claim. Density dependence makes $Q$ state-dependent; changing $G_t$, $M_t$, or the score parameters makes it time-dependent. Either change invalidates the fixed-matrix proof. Potential games can converge under other hypotheses, while frequency-dependent replicator systems can cycle or behave chaotically \citep{hofbauer1998evolutionary,galla2013complex}. A nonautonomous system can also have a moving instantaneous Perron vector that the population fails to track. ROM supplies no theorem that selects among those cases without a specified system, a solution concept, and conditions appropriate to that system.

\subsection{Correspondences and their boundaries}

Equation~\eqref{eq:rom-main} sits within an established selection--transmission family \citep{hadeler1981stable,page2002unifying}. Three correspondences help interpret pieces of the template, but none supplies missing empirical or asymptotic content.

\paragraph{Price partition.} In an appropriate discrete-time representation, change in a trait mean can be partitioned into selection and transmission terms,
\[
\Delta\bar z=\frac{1}{\bar w}\operatorname{Cov}(w,z)
+\frac{1}{\bar w}\mathbb E[w\Delta z].
\]
This is an accounting correspondence, not an independent evolution law for $w$, $\rho$, $M$, or $G$.

\begin{remark}[Bayesian update correspondence]
\label{rmk:bayesian}
With identity transfer in normalized discrete time, $p_{t+1}(\tau)\propto p_t(\tau)w(\tau)$ has the algebra of a Bayesian update when $w$ is assigned the role of a likelihood \citep{harper2009replicator,czegel2022bayes}. The shared normalization rule does not imply that the full continuous-time, state-dependent ROM equation is a Bayesian filter. It also does not imply mutual-information maximization, average-log-fitness optimality, or convergence. Those claims require a joint environment model, prediction target, and objective.
\end{remark}

\paragraph{Reinforcement learning.} Some softmax policy-gradient dynamics take a replicator form with action values in the payoff position \citep{tuyls2003selection,bloembergen2015evolutionary}. In a specified learning model, $M$ may represent exploration or imitation. Independent Q-learning and value decomposition do not thereby instantiate Equation~\eqref{eq:rom-main}; the companion MARL study is a construct test of the proposed index, not a simulation of institutional ROM dynamics.

Pure-selection replicator dynamics possess a Shahshahani-gradient representation when the fitness field is generated by a potential. Appendix~\ref{app:gradient-flow} states the restricted theorem and a non-potential counterexample. Mutation, frequency dependence, and time dependence do not inherit that potential automatically.

\subsection{The kernel triple and ownership-sensitive transfer}
\label{sec:kernel-triple}

The modeler's central declaration is the triple
\begin{equation}
\mathcal K_S=(\rho_S,w_S,M_S).
\label{eq:kernel-triple}
\end{equation}
The first two components enter observable frequency dynamics through their product, while $M_S$ directs the resulting mass across destinations. Their interpretation is application-specific: genetic mutation, imitation, organizational reform, or policy diffusion are not interchangeable mechanisms simply because each can be represented by a stochastic matrix.

This triple is a bookkeeping device for causal commitments. Moving a covariate from $\rho$ to $w$ leaves the vector field unchanged if their product is preserved, but it changes the intended intervention and identification story. Moving it into $M$ changes something stronger: the destination of propagated mass rather than only the source's total contribution. Applications should therefore report both the numerical kernel and the semantic rule used to estimate each row. A fitted transition matrix without a stable type coding can confound genuine transfer with changes in classification.

One candidate transfer mechanism tilts a baseline kernel $M^0$ toward destinations with larger scalar ownership score $O$:
\begin{equation}
M_S(i\to j)=\frac{M^0_{ij}e^{\gamma O_j}}{\sum_kM^0_{ik}e^{\gamma O_k}}.
\label{eq:ownership-kernel}
\end{equation}
The row normalization is essential. The apparent source factor in $e^{-\gamma(O_i-O_j)}$ cancels, leaving a destination tilt. If $M^0$ is reversible, this scalar tilt remains reversible with a reweighted stationary measure; it does not generically create circulation or cycles. Appendix~\ref{app:gradient-flow} proves the identity, and Appendix~\ref{app:ownership-microfoundations} states the narrow exit-probability comparative static and empirical tests.

Reversibility is a statement about pairwise stationary flows, not simply about symmetry of the displayed matrix. A kernel is reversible with respect to $\pi$ when $\pi_iM_{ij}=\pi_jM_{ji}$ for every pair. A nonreversible kernel may nonetheless have a unique stationary mass distribution; the difference is that its stationary edge currents need not vanish. Appendix~\ref{app:gradient-flow} gives an explicit three-state chain with convergent mass and a positive stationary current. That current records circulation of transitions through edges. It is not a periodic orbit of $p_t$, and it does not show that an institutional population cycles.

\subsubsection{Four distinct long-run statements}

The paper uses four notions that should not be collapsed. First, a \emph{stationary distribution of a kernel} satisfies $\pi M=\pi$. Second, \emph{detailed balance} is the stronger pairwise condition $\pi_iM_{ij}=\pi_jM_{ji}$; it implies stationarity but not conversely. Third, a \emph{stationary current}
\begin{equation}
J_{ij}:=\pi_iM_{ij}-\pi_jM_{ji}
\label{eq:stationary-current-definition}
\end{equation}
can be nonzero while $\pi$ itself remains constant because incoming and outgoing currents balance at every node. Fourth, a \emph{population limit cycle} is a nonconstant periodic orbit of the autonomous vector field for $p$. It is a property of trajectories through the simplex, not of edge counts at a stationary composition.

These distinctions block two tempting but invalid inferences. Breaking detailed balance does not prove a population cycle, and proving kernel reversibility does not prove convergence of a state-dependent weighted ROM system. The normalized ownership tilt yields an exact detailed-balance result only when its baseline kernel is reversible. The static Perron theorem yields continuous-time composition convergence under a fixed irreducible weighted matrix whether or not the associated transfer kernel is reversible. Frequency-dependent payoffs can create a non-potential vector field even with identity transfer. The relevant long-run statement must therefore be attached to the operator for which it was proved.

Time variation adds a fifth case rather than a shortcut. If $M_t$ or $\phi_t$ changes, each frozen-time operator may have a well-defined stationary or Perron vector, yet $p_t$ need not equal or track that moving vector. Tracking requires a quantitative separation between the rate of environmental change and the contraction or spectral gap of the population dynamics. No such bound is part of ROM. Describing a sequence of instantaneous solutions as a ``moving equilibrium'' is therefore definitional unless an independent tracking theorem is supplied.

\subsection{Coarse-graining and exact closure}
\label{sec:coarse-graining}

Let $\pi:T_S\to T_{S'}$ partition the fine types. A single first-order Markov kernel on the coarse types reproduces the projected transfer dynamics for every initial fine distribution if and only if the fine kernel is strongly lumpable with respect to $\pi$ \citep{kemeny1976finite,geiger2022information}.

\begin{theorem}[Strong-lumpability criterion]
\label{thm:lumpability}
For a finite row-stochastic kernel $M_S$, universal first-order closure under $\pi$ holds if and only if, for every pair $x,x'$ in the same source block and every destination block $B$,
\begin{equation}
\sum_{y:\pi(y)=B}M_S(x,y)
=\sum_{y:\pi(y)=B}M_S(x',y).
\label{eq:lumpability-condition}
\end{equation}
For Equation~\eqref{eq:rom-main}, constancy of effective fitness $\phi=w\rho$ within each source block is an additional sufficient condition for closure in the same weighted form. Kernel lumpability alone does not generally remove dependence on hidden within-block composition. Separate coarse interpretations of $w$ and $\rho$ require separate restrictions; the product alone does not identify them.
\end{theorem}

When Equation~\eqref{eq:lumpability-condition} holds, the common block-transition sum defines the quotient kernel
\begin{equation}
K(A,B):=\sum_{y:\pi(y)=B}M(x,y),
\qquad x\in\pi^{-1}(A).
\label{eq:quotient-kernel}
\end{equation}
The condition makes the definition independent of the representative $x$. If $\phi(x)=\Phi(A)$ for $x\in\pi^{-1}(A)$ and $P(A)=\sum_{\pi(x)=A}p(x)$, summing the fine equation over a destination block gives
\begin{equation}
\dot P(B)=\sum_A P(A)\Phi(A)K(A,B)
-P(B)\sum_A P(A)\Phi(A),
\label{eq:weighted-quotient}
\end{equation}
which is again ROM at the coarse scale. This derivation is exact; no averaging convention for hidden fitness is needed because the within-block values coincide.

When lumpability fails, two point masses in the same coarse block generate different next-step coarse distributions, so no one coarse kernel works for every fine initial state. This proves failure of universal first-order closure. It does not prove that every fitted coarse model is inaccurate, select a unique memory kernel, or determine the sign of approximation error. A kernel fitted under one stationary ensemble can work for that ensemble while failing after an intervention changes the within-block mixture. This is precisely why ``works on historical frequencies'' is weaker than closure for every initial distribution.

Hidden fitness variation creates an analogous problem even with a lumpable transfer kernel. Two fine distributions can have the same block masses but different fitness-weighted source totals. Unless those totals are functions of the block masses under the admissible ensemble, their coarse derivatives differ. One can enlarge the retained state with conditional within-block moments, use a history-dependent model, or fit an approximate closure. Each option changes the estimand; none is automatically supplied by the projection.

For nested projections $T_0\to T_1\to T_2$, exact lumpability at each step implies that the direct quotient equals the stepwise quotient at every time. Outside exact closure, direct and stepwise approximations can use different estimators, state spaces, horizons, and losses; no general error ordering follows. Appendix~\ref{app:ladder-constraint} gives the theorem and proof. Time-scale separation, mean-field limits, symmetry, or controlled renormalization can support an approximation only with their own assumptions and error analysis \citep{mori1965transport,zwanzig1961memory,aristoff2023coarsegraining,villegas2023laplacian}.

Composite lumpability does not require a chosen intermediate partition to be lumpable. It is therefore possible for a direct coarse description to close even when one proposed ladder does not. Conversely, good adjacent fits do not imply a controlled composite approximation unless the error propagation is analyzed. These two facts rule out both universal slogans: neither ``never skip a level'' nor ``coarsen directly'' follows from the theorem. The stable conclusion is the intertwining identity proved in Appendix~\ref{app:ladder-constraint}; everything outside it is model comparison.

Exact closure can be time-indexed without becoming stationary. If the same partition satisfies the block-sum condition for $M_t$ at every time and $\phi_t$ factors through its blocks, the summation leading to Equation~\eqref{eq:weighted-quotient} yields an exact nonautonomous quotient. The quotient parameters then inherit time dependence, so Theorem~\ref{thm:rom-stationary} still does not apply. Closure asks whether retained variables suffice for projected evolution; stationarity asks about the long-run behavior of a fixed autonomous operator. Neither answer supplies the other.

Interventions impose a stronger transport question. A partition can be lumpable for the observed kernel but cease to be lumpable after an intervention changes transitions differently inside a block. Likewise, observationally constant block fitness can split under a targeted treatment. A coarse causal model should therefore require the relevant family of intervened kernels and fitness maps to factor through the retained state, not only the historical operator. Observational closure does not guarantee interventional closure.

Effective information and related multiscale statistics can help compare specified descriptions, but they do not choose a scale by theorem. A defensible comparison estimates micro and macro transition models from commensurable observations or interventions and evaluates both on the same task. Higher macro-level effective information without out-of-sample improvement is not evidence of ontological causal superiority.

\section{Conditional institutional instantiation}
\label{sec:institutional}

The institutional example maps ROM's abstract objects to governance measurements. It is an instantiation, not a consequence of the general equation. The units are affected parties and decision procedures: a party bears consequences; a controller or procedure selects among alternatives. Calling a controller a consent-holder requires a separate authorization rule. The model does not infer authorization from control, influence, or survival.

\subsection{Friction as a proposed index}

Let $s_i(d)\geq0$ be affected party $i$'s stake in domain $d$, divided by a declared reference stake $s_{0,S}>0$ to give $\tilde s_i=s_i/s_{0,S}$. Let $\alpha_i\in(-1,1]$ be signed preference--decision alignment and $\varepsilon_i\geq0$ a declared information-loss coordinate. The proposed dimensionless index is
\begin{equation}
F(d,t)=\sum_i\tilde s_i(d)
\frac{1+\varepsilon_i(d,t)}{1+\alpha_i(d,t)}.
\label{eq:friction}
\end{equation}
Under homogeneous coordinates it becomes $F=\sigma(1+\varepsilon)/(1+\alpha)$ with $\sigma=\sum_i\tilde s_i$.

Each coordinate needs a unit of observation. Stakes attach to an affected party, a domain, and a consequence window. Multiplying every raw stake and $s_{0,S}$ by the same positive constant leaves $\tilde s_i$ and $F$ unchanged, so ordinary unit conversion is harmless. Replacing only the benchmark by $s'_{0,S}=c\,s_{0,S}$ instead sends $F$ to $F/c$ and the stipulated bridge below to $L/(1+F/c)$; because of the additive one, that is a nonlinear change in $\rho$, not a gauge symmetry. An application must therefore fix $s_{0,S}$ ex ante through a substantive calibration tied to a stated consequence unit, horizon, and comparison population, must not tune it to the persistence outcome, and must report sensitivity to defensible alternatives. Absolute $\rho$ values from differently calibrated domains are not commensurable without an external linking rule. Alignment attaches a reported or elicited preference to a particular decision under a signed coding. It is not general ideological agreement, similarity between two agents, or correlation magnitude. Information loss attaches to an observation channel and estimator. These indices can vary within the same institution across decisions, so an institutional type needs a declared aggregation rule before it receives one value of $F$.

The affected-party set is also part of the measurement. Controllers, beneficiaries, bystanders, and persons exposed to downside need not coincide. Including only formal voters can convert exclusion into missing data; including every remotely affected person can make stakes impossible to estimate. The model supplies no automatic boundary. A defensible application fixes an inclusion rule, records contested cases, and tests whether conclusions survive plausible changes to the set and admissible benchmark calibration.

Equation~\eqref{eq:friction} is not itself stake--voice mismatch: voice does not appear in it. Stakes weight the index, alignment links preferences to decisions, and $\varepsilon$ represents information loss under a separate measurement model. Effective voice enters the survival specification below. The positive floor $F=\sigma/2$ at $\alpha=1,\varepsilon=0$ and the pole as $\alpha\downarrow-1$ are properties of the chosen normalization, not measured laws or instability theorems. Appendices~\ref{app:lagrangian} and \ref{app:information-motivation} show why constrained optimization and information theory do not uniquely derive the ratio.

\begin{remark}[Quadratic refinement withdrawn]
\label{rmk:quadratic-refinement}
An earlier exploratory fit favored a squared target-correlation form, but its generator made positive and negative treatment labels share the same correlation magnitude. The apparent symmetry was built into the treatment. The quadratic form is retained only as a record of functional-form underdetermination, not as evidence \citep{farzulla2026marl}.
\end{remark}

\subsection{Effective voice and conditional survival}

Define the descriptive effective-voice score
\begin{equation}
L(d)=\frac{\sum_i s_i(d)v_i(d)}{\sum_i s_i(d)},
\qquad 0\leq v_i(d)\leq1,\qquad \sum_i s_i(d)>0,
\label{eq:legitimacy}
\end{equation}
where $v_i$ is measured influence or control available to affected party $i$. The historical symbol $L$ is retained for compatibility with the formalization, but in this paper it denotes effective voice. It is not political legitimacy. Calling $L$ a legitimacy measure requires the AOC's normative authorization premise plus a defended measurement bridge; proportional influence, expressed preference, and valid consent are not equivalent by definition \citep{farzulla2025aoc,farzulla2025consensual}.

The level interpretation of $v_i$ is load-bearing. If every affected party has almost no influence, proportional \emph{shares} of that influence can still look perfectly matched to stake shares. Equation~\eqref{eq:legitimacy} instead averages absolute influence levels using stakes as weights, so collective disenfranchisement remains low effective voice. This does not make the score a complete theory of authorization: agenda control, option quality, capacity, disclosure, voluntariness, and the right to refuse can all be omitted while $v_i$ is high.

The conditional ROM instantiation sets
\begin{equation}
\rho_S(\tau)=\frac{L(\tau)}{1+F(\tau)},
\qquad 0\leq L\leq1,\quad F\geq0.
\label{eq:effective-voice-survival}
\end{equation}
Equation~\eqref{eq:effective-voice-survival} is a candidate type-level specification only after decision-, domain-, and time-aggregation rules for $F$ and $L$ have been prespecified; without them the application has not instantiated $F(\tau)$, $L(\tau)$, or the resulting effective fitness $\phi(\tau)$. At fixed $w$, $M$, and other covariates, the equation rises with measured effective voice and falls with the stipulated friction index. These are algebraic comparative statics. Until a persistence event, horizon, and outcome link are declared, $\rho_S(\tau)$ is only a candidate model score, not an estimated survival probability or hazard. Whether the score predicts institutional duration, replacement, or another outcome is an empirical question; whether higher effective voice is legitimate is a normative question.

Under the static theorem's restrictions, $L$ and $F$ enter stationary composition only through $\phi=wL/(1+F)$. Even there, the componentwise ranking established here follows under the exact uniform-residual specification; general kernels need not preserve it and can give a lower-fitness type more stationary mass through inflow. Thus ROM does not establish that history selects for consent or that lower-friction institutions necessarily prevail.

\subsection{Observed and latent contention}

Observed protest, litigation, noncompliance, exit, or volatility are outcomes, not the definition of $F$. Low observed contention can coexist with preference--decision divergence when expression or exit is costly. Conversely, visible disagreement can occur under legitimate authorization. A study must therefore distinguish the proposed latent score from its observed indicators and state an outcome link such as
\begin{equation}
\mathbb E[Y_{d,t}\mid F_{d,t},L_{d,t},X_{d,t}]=g(F_{d,t},L_{d,t},X_{d,t}).
\label{eq:outcome-link}
\end{equation}
The function, controls, timing, and measurement errors must be fixed before inference. Suppression capacity and information loss are distinct constructs: censorship may affect a particular information measure, but no value of $\varepsilon$ or latent $F$ follows from the label alone. Detailed suppression dynamics belong to an application model, not to ROM's general equation.

The normative bridge is equally explicit. ROM can compare conditional persistence after an application defines $L$, $F$, $w$, and $M$. It cannot infer that persistent arrangements deserve obedience, that affected parties endorse them, or that lowering the index is good. A conditional instrumental claim requires an independently defended premise about affected parties' interests or authorization. The AOC supplies one proposed premise; ROM neither proves nor presupposes its correctness.

\section{Measurement, identification, and evidence}
\label{sec:operational}

\subsection{A measurement protocol}

A testable instantiation begins with distinct measurement objects:
\begin{itemize}[noitemsep]
    \item normalized stakes $\sigma$, including the unit and reference stake;
    \item signed preference--decision alignment $\alpha$, including whose preferences and which decisions;
    \item information loss $\varepsilon$, including the estimator and observation channel;
    \item effective voice $L$, including the level of influence rather than voice shares alone;
    \item baseline capacity $w$ and transfer kernel $M$; and
    \item an observed outcome $Y$ that is not defined to equal the modeled index.
\end{itemize}
The protocol should prespecify the outcome link, covariates, horizon, support, missingness rules, and comparison models. The canonical ratio earns empirical support only if it improves held-out discrimination over stakes-only, additive, interaction-rich, and flexible alternatives. Its algebraic monotonicity is not evidence that the measured world follows it.

Falsification is local to the instantiation. Failure of $F$ to predict its declared outcome defeats that measurement and functional form in the tested setting; it does not falsify the generic row-stochastic identity. Conversely, a good fit does not validate the normative interpretation of $L$ or transport the result to another environment.

\subsection{Identifiability of baseline weight and survival}
\label{sec:identifiability}

The ROM equation contains baseline weight $w_S$ and survival factor $\rho_S$ only through their product
\[
\phi_S(\tau,G,p,t)=w_S(\tau,t)\rho_S(\tau,G,p,t).
\]
Even with the transfer kernel known, observed frequency changes identify this effective-fitness product only under additional dynamical and measurement assumptions. They do not uniquely separate its factors.

The within-model-class non-identification is exact. For any positive admissible function $h_S(\tau,t)$ whose arguments preserve the declared baseline-rate class, define
\begin{equation}
w'_S(\tau,t)=w_S(\tau,t)h_S(\tau,t),
\qquad
\rho'_S(\tau,G,p,t)=\frac{\rho_S(\tau,G,p,t)}{h_S(\tau,t)}.
\label{eq:identification-transformation}
\end{equation}
Where the transformed factors remain in their declared domains, $w'_S\rho'_S=w_S\rho_S$, so the ROM vector field is unchanged without changing either factor's argument domain. Constant rescaling is only one member of this observationally equivalent family. A broader positive reallocation $h_S(\tau,G,p,t)$ can also preserve the product algebraically, but then $w'_S=w_Sh_S$ generally depends on $G$ or $p$ and leaves the declared baseline class $w_S(\tau,t)$. That is an equivalence in an enlarged parameterization, not a gauge transformation within the stated effective-fitness specification $\phi_S=w_S\rho_S$.

The domain qualification is important. If an application declares $0\leq\rho\leq1$, a proposed $h_S(\tau,t)$ is admissible only when $\rho/h_S$ remains in that interval and the transformed $w$ remains a valid rate with the same argument domain. Those restrictions can shrink the equivalence class, but they do not select a unique point within it. Even a perfectly observed continuous path $p_t$ cannot distinguish two admissible decompositions whose products agree at every visited state. Data away from the visited path, an intervention, or a structural restriction is needed to do more.

This symmetry can be read as a parameterization gauge: frequency dynamics depend on $\phi$, while the names ``capacity'' and ``survival'' allocate that product between two conceptual channels. The gauge is harmless for forecasting $p$ when $\phi$ is stable, but it is decisive for intervention. A policy said to raise $w$ and one said to raise $\rho$ can have the same fitted trajectory yet imply different measurements, side effects, and transport assumptions. ROM therefore requires the factor labels to be defended by variation external to the product rather than by fit alone.

Separate identification therefore requires information outside the frequency path. Three common routes are:
\begin{enumerate}[noitemsep]
    \item a normalization fixing one factor's scale;
    \item an exclusion restriction supplying variation that changes one factor without directly changing the other; or
    \item a structural measurement model whose factors have distinct observable implications.
\end{enumerate}
Repeated environments can help only when their invariances are stated. For example, one might assume that a measured resource shifter changes $w$ while leaving the survival mapping fixed, or that an authorization intervention changes a measured input to $\rho$ while leaving baseline capacity fixed. Such exclusions are scientific assumptions open to failure. If the resource intervention also changes participation, information, or transfer, the proposed separation no longer follows.

The kernel introduces a second identification layer. Equation~\eqref{eq:rom-main} observes the fitness-weighted inflow $p\operatorname{diag}(\phi)M$, not every entry of $M$ in isolation. Transition microdata with source and destination labels can identify rows under sampling and stationarity assumptions; aggregate type shares generally cannot. When both $\phi$ and $M$ are unknown, changes in apparent source success may be traded against destination flow over a limited trajectory. A full-rank design, direct transition observations, or restrictions on $M$ are therefore needed before interpreting a fitted entry as imitation, reform, or replacement.
The consent-friction instantiation sets $\rho_S=L/(1+F)$, but that parameterization does not identify $L$, $F$, or $w_S$. Each still requires declared measurements, timing, and variation. Labels such as ``resource shock'' and ``authorization shock'' do not establish exclusion restrictions by themselves, because resources can alter effective voice and governance responses can alter resources.

Duration and competing-risk designs can contribute evidence only after event types and hazards are linked to the model. For example, a study may posit separate resource- and governance-related exit hazards, but their additivity and their relation to $w_S$ and $\rho_S$ are extra assumptions rather than consequences of Equation~\eqref{eq:rom-main}. The same caution applies to cross-sectional interaction designs: complementarity between resources and effective voice must be estimated, not inferred from multiplicative non-identification.

Time units create a simpler ambiguity. Multiplying every effective-fitness rate by a common positive constant rescales the speed of an autonomous trajectory without changing its path through the simplex. Observations with calibrated clock time can identify that speed; an unordered sequence of compositions cannot. This is distinct from Equation~\eqref{eq:identification-transformation}, which leaves even the time-indexed vector field unchanged. Both should be separated from ordinary sampling uncertainty.

A useful identification report therefore lists four objects: the equivalence class left by the observed path; the normalization that chooses a representative; the data or intervention intended to separate channels; and the sensitivity of the substantive conclusion to alternative admissible representatives. If only $\phi$ is identified, the honest estimand is effective fitness. The paper should not relabel its factors as separately measured causes.

This limitation is load-bearing. ROM can predict relative persistence from a prespecified $\phi$; it cannot infer that persistence was caused by capacity, effective voice, low friction, or any particular decomposition without independent measurements or restrictions. The superseded longer catalogue of candidate designs and political examples remains as an excluded archival source file; it is not included in either PDF because its causal examples require reconstruction.

\subsection{Evidence status}
\label{sec:evidence-status}

The original factorial is retained only as a design-failure record: its learner arms had unequal budgets, its target-correlation generator was sign-blind, observation noise lacked a causal path in one arm, and the resulting squared fit is withdrawn. The retained repaired IQL and VDN bundles each use the same environmental budget of 1{,}000 100-step episodes per replication and matched problem draws; this does not equate optimizer updates or restore the auxiliary MAPPO arm to headline status. Per resource, the signed sampler draws four-agent targets from $\mathcal N(\mu,(1-\rho_{\mathrm{tar}})I+\rho_{\mathrm{tar}}\mathbf1\mathbf1^\top)$, verifies the realized signed correlations before training, and uses translated feasible-centred support rather than clipping in the support control. Observation noise enters each agent's state observation directly during training; clean/noisy problems and initializations are paired, trained policies are frozen, and the signed-noise crossing is scored on common held-out resets under clean, matched-noise, and restored-exploration protocols.

The current mixed-motive control battery reports exploratory evidence against the implemented proxy-ratio instantiation in its four-agent resource environment while retaining a scoped shared-state signed target-coordinate-correlation effect. Here $\rho_{\mathrm{tar}}$ is correlation among agents' ideal points inside a fixed quadratic reward family, not a direct manipulation or measure of objective- or reward-function alignment. Under full separation, IQL is structurally invariant to cross-agent target-coordinate correlation and VDN is a non-detection, not an equivalence result. Paired legacy-support partial sharing shows overlap-related modulation but does not establish a universal dose law; feasible-centred held-out evaluation detects opposition penalties and an exact finite-horizon oracle splits the positive-target-correlation gradient between feasibility and a support-dependent frozen-policy residual. In a reduced feasible-centred frozen-policy signed-noise crossing, the mean noise direction is positive but the target-correlation--noise interaction runs opposite to the proxy ratio's prediction. These are exploratory controlled demonstrations, not institutional ROM dynamics: they establish neither convergence, authorization, nor a general law of information loss. Exact coefficients, intervals, filters, and support qualifications are held by the owning MARL paper \citep{farzulla2026marl}, whose code, data, inventories, and validation record are public as Zenodo v3.0.0, \href{https://doi.org/10.5281/zenodo.22004215}{doi:10.5281/zenodo.22004215}; that paper's claim map and technical supplement are not yet public.

\section{Worked example: medical delegation}
\label{sec:example}

Consider a patient affected by a treatment choice and a physician controlling the immediate decision. The example illustrates the separation of effective voice from preference--decision alignment; it does not determine the ethically correct treatment.

Let normalized stakes be $\sigma=1$, information loss $\varepsilon=.4$, and measured effective voice $L\in[0,1]$. Table~\ref{tab:medical-example} applies the homogeneous index and survival score.
\begin{table}[htbp]
\centering
\caption{Illustrative coordinates for one delegated decision. Values are assigned to expose distinctions, not estimated clinical effects.}
\label{tab:medical-example}
\begin{tabular}{lcccc}
\toprule
Scenario & $\alpha$ & $L$ & $F=(1+\varepsilon)/(1+\alpha)$ & $\rho=L/(1+F)$\\
\midrule
Low voice, moderate alignment & $.6$ & $.1$ & $.875$ & $.053$\\
Shared decision, moderate alignment & $.6$ & $.5$ & $.875$ & $.267$\\
High voice, weak alignment & $.2$ & $.8$ & $1.167$ & $.369$\\
\bottomrule
\end{tabular}
\end{table}

The first two rows hold alignment and information fixed while changing measured influence. The index $F$ is unchanged because voice is not one of its arguments; the conditional survival score rises because $L$ rises. The third row has high measured influence but lower preference--decision alignment, so the assigned $F$ is larger even though $L$ remains high. Effective voice and friction are therefore distinct coordinates.

No clinical consequence follows without Equation~\eqref{eq:outcome-link}. Adherence, satisfaction, health, and complaints are different outcomes and can move in different directions. Nor does $L=.8$ establish valid informed consent: capacity, disclosure, voluntariness, and authorization require a normative and clinical measurement framework outside ROM. The example's purpose is only to make the conditional model auditable.

\section{Discussion}
\label{sec:discussion}

\subsection{What ROM contributes}

ROM's contribution is organizational rather than a new population equation. Its value lies in forcing a modeler to expose the type space, effective-fitness decomposition, transfer process, environment, and scale map before interpreting persistence. The restricted stationary theorem and lumpability criterion then show which conclusions follow after strong conditions are imposed. The framework earns explanatory value only when an application supplies measurements and predicts better than simpler alternatives.

This position is narrower than generalized Darwinism, which already argues that variation, selection, and retention recur in social and economic evolution \citep{hodgson2010darwin}. It is also narrower than the Price equation, which supplies a substrate-neutral accounting identity. ROM's added structure is the scale-indexed triple $(\rho,w,M)$ plus an explicit closure and evidence ledger. Whether that structure improves an empirical analysis remains open.

The propensity interpretation of fitness is load-bearing. Effective fitness refers to an expected rate under the stipulated environment, not an after-the-fact label for whatever survived \citep{millsbeatty1979}. This makes the model falsifiable: a prespecified $\phi$ can fail to predict relative persistence. It also prevents the word ``optimization'' from becoming a tautology.

\subsection{Adjacent formalisms}

Mori--Zwanzig projection gives one formal route from omitted variables to memory terms, but ROM does not derive a projection operator for institutions \citep{mori1965transport,zwanzig1961memory}. Renormalization theory supplies precise scale maps in specified physical systems; the ROM ladder is instead the standard finite-state lumpability criterion plus application-level approximation questions. The Price equation partitions change but does not specify the laws governing $G$, $w$, $\rho$, or $M$.

The Free Energy Principle is a separate persistence framework whose variational free energy is not Equation~\eqref{eq:friction} \citep{friston2010free}. Appendix~\ref{app:gradient-flow} establishes a potential only for a pure-selection slice with an integrable fitness field. ROM offers no general free-energy, entropy-production, or convergence principle for its state-dependent form.

The ownership kernel also illustrates why exact normalization matters. A scalar destination tilt of a reversible baseline stays reversible. A stationary probability current can exist in a nonreversible chain while its mass distribution converges. Neither a tilted kernel nor a nonzero current is evidence of a population limit cycle. Long-run behavior must be proved for the specified autonomous system.

\subsection{Limitations and boundary conditions}

The deterministic equation is least secure in small or strongly dependent populations, where stochastic drift and finite-sample variation can dominate. The finite type space excludes continuous and evolving representations. Network position can alter both effective fitness and transfer; Appendix~\ref{app:network-topology} gives diagnostics, not a universal correction. Parameters may also co-evolve with the population. Unless their evolution laws are written, ROM treats them as exogenous paths rather than endogenous mechanisms.

Identification remains a central obstacle. The population path identifies $w\rho$ more readily than either factor, and the institutional variables add measurement error and contested construct boundaries. Effective voice is not authorization, observed calm is not low latent contention, and a high persistence rate is not legitimacy. These are not verbal caveats to be relaxed after fitting; they determine what the model means.

The empirical record is also limited. The MARL battery studies a small quadratic resource game, two learner/objective bundles, and exploratory controls. Its contradictions matter for the implemented proxy-ratio instantiation because that index made directional and single-ratio predictions in the tested environment. They do not show that every possible target-correlation treatment, information channel, or institutional application behaves similarly, still less that target correlation is objective alignment. Independent replication and application-specific data remain absent.

Finally, the formal verification boundary is narrow. Lean checks algebraic identities and monotonicity of defined scalar functions. It does not certify the stationary theorem, existence of ODE solutions, nonnegative invariance, coarse-graining approximation, empirical estimands, or the ethical bridge. Appendix~\ref{sec:appendix-lean} maps each checked theorem to its exact content.

\subsection{Descriptive and normative separation}

ROM's supported position is instrumental. Given a declared outcome and independently defensible measurements, it can ask whether a configuration's effective-fitness specification predicts relative persistence. It cannot infer objective value from that persistence. A claim that lower friction is preferable requires a premise about affected parties' interests, rights, or authorization. A claim that high effective voice is legitimate requires more than an influence score. The AOC proposes such a bridge and faces the usual challenge that an evolutionary origin of a preference may explain rather than justify it \citep{street2006darwinian}. ROM leaves that dispute external.

This separation also disciplines intervention claims. Raising measured voice can change $L$ without changing $F$; increasing transparency can change an information estimator without changing alignment; altering resources can change $w$ and the feasible set simultaneously. A policy conclusion requires a causal design for those channels and an independent normative criterion. The reader paper and technical supplement omit the earlier policy, finance, DAO, and prospective-application catalogue rather than presenting it as a consequence of the dynamics. It remains only in the immutable long-record archive and is not part of the reader evidence package.

\section{Conclusion}
\label{sec:conclusion}

ROM is a finite-type persistence template, not a cross-domain law. It separates baseline weight, conditional survival, and transfer at a declared scale; preserves total mass under row-stochastic transfer; and converges in continuous time only under the finite, static, density-independent positive-fitness and irreducible-weighted-kernel restrictions of Theorem~\ref{thm:rom-stationary}. Discrete-time power convergence requires the stronger primitive case. Strong lumpability characterizes universal exact closure of the transfer kernel, while blockwise effective fitness supplies an exact weighted-ROM quotient. The componentwise stationary ranking is guaranteed here under uniform residual transfer, and the general state-dependent system has no convergence or potential theorem here.

The institutional instantiation is correspondingly conditional. Its $L$ is descriptive effective voice, its friction ratio is an empirically rejectable index, and political legitimacy requires an external normative and measurement bridge. The MARL battery retains a shared-state signed target-coordinate-correlation effect but reports exploratory evidence against the implemented proxy-ratio instantiation and, in a reduced feasible-centred frozen-policy crossing, reverses its predicted target-correlation--noise interaction. It does not identify objective- or reward-function alignment. The useful result is therefore an assumptions ledger: state what is defined, identify what is proved, and let unsupported mappings fail visibly.

\section*{Code and artifact availability}

Publicly archived code and control-battery artifacts are available in the cited repositories \citep{farzulla2026marl,farzulla2026lean}. The companion's Zenodo v3.0.0 release inventory and checksums identify the evidence state supporting the current paragraph; earlier repository and archive states do not contain the complete follow-up package. This 37-page reader edition follows the explicit 19 August 2026 waiver of the secondary 51--54-page aspiration. The checksum-identified long technical record and the standalone technical supplement remain preserved separately.

\section*{Declarations}

\paragraph{Conflict of Interest.} The author declares no competing interests.

\paragraph{Funding.} This research received no external funding.

\paragraph{Data Availability.} The computational evidence is described in Section~\ref{sec:evidence-status}. Reproduction uses the synchronized companion release inventory and checksums at \href{https://doi.org/10.5281/zenodo.22004215}{doi:10.5281/zenodo.22004215}; repository and archive states predating the August repair are insufficient. That record is a dataset release (code and data archives, inventories, and a validation report); the companion manuscript's claim map and technical supplement are not yet public.

\paragraph{AI Assistance.} Claude supported formal verification, literature synthesis, adversarial cross-review, and earlier LaTeX preparation. OpenAI Codex supported source review, compression, LaTeX editing, and build and visual quality assurance for this reader-facing working draft. The author reviewed all claims and takes sole responsibility for the final text.

\appendix

\section{Friction-index motivations and non-identification}
\subsection{Why constrained optimization does not identify the friction ratio}
\label{app:lagrangian}

The canonical index
\[
F=\sigma\frac{1+\varepsilon}{1+\alpha}
\]
can be embedded in a constrained allocation problem, but that embedding does not derive the ratio. This appendix keeps the useful negative result: a natural linear program both assumes the target multiplier and produces a corner allocation. The exercise is therefore an internal-consistency check and a warning against describing the index as an optimum.

Let affected party $i$ have dimensionless stake $s_i\geq0$, assigned effective voice $v_i\in[0,1]$, signed preference--decision alignment $\alpha_i\in(-1,1]$, and information-loss coordinate $\varepsilon_i\geq0$. Consider the stipulated individual cost
\begin{equation}
c_i(v_i,\alpha_i,\varepsilon_i)
=\frac{(1-v_i)(1+\varepsilon_i)}{1+\alpha_i},
\label{eq:individual-cost}
\end{equation}
and the fixed-capacity program
\begin{equation}
\min_{v}\ \sum_i s_i c_i(v_i,\alpha_i,\varepsilon_i)
\quad\text{subject to}\quad
0\leq v_i\leq1,\qquad \sum_i v_i=V_0,\qquad 0\leq V_0\leq n.
\label{eq:voice-program}
\end{equation}
Equation~\eqref{eq:individual-cost} already contains $(1+\varepsilon_i)/(1+\alpha_i)$. Monotonicity in information loss and alignment does not select that ratio: bounded logistic, additive, saturating, and interaction-rich alternatives can satisfy the same qualitative desiderata. Optimization cannot independently validate a multiplier supplied in the objective.

The ratio also imposes more than direction. Its marginal effects are
\[
\frac{\partial F}{\partial \varepsilon}
=\frac{\sigma}{1+\alpha},
\qquad
\frac{\partial F}{\partial \alpha}
=-\frac{\sigma(1+\varepsilon)}{(1+\alpha)^2},
\qquad
\frac{\partial^2F}{\partial\alpha\,\partial\varepsilon}
=-\frac{\sigma}{(1+\alpha)^2}.
\]
Thus it predicts an alignment-dependent noise penalty that is largest toward opposition, a positive floor at maximal alignment, and divergence at the lower alignment boundary. These are discriminating restrictions, not consequences of saying that more stakes or information loss should increase a cost. In the signed control battery's proxy treatment, the estimated noise penalty instead grows toward more positive target-coordinate correlation.

The Lagrangian first-order condition for an interior coordinate is
\begin{equation}
\lambda=\frac{s_i(1+\varepsilon_i)}{1+\alpha_i}.
\label{eq:voice-foc}
\end{equation}
Because the objective is linear in every $v_i$, its marginal is constant in $v_i$. Order parties by the displayed marginal reduction, allocate voice to the largest, and saturate each upper bound before moving to the next. Absent ties, every coordinate is at zero or one except possibly one fractional coordinate when $V_0$ is not an integer. A proportional interior allocation is not an output of this program. It would require additional curvature, such as a strictly convex voice cost, and even then would not generally be proportional to stakes.

For homogeneous $\alpha_i=\alpha$ and $\varepsilon_i=\varepsilon$, aggregate stipulated cost is
\begin{equation}
F^*=\Delta\frac{1+\varepsilon}{1+\alpha},
\qquad
\Delta:=\sum_i s_i(1-v_i^*).
\label{eq:total-friction}
\end{equation}
The optimization therefore yields a family indexed by the residual voice deficit $\Delta$, not the aggregate-stakes index used in the main text. Substituting $\Delta=\sigma:=\sum_i s_i$ gives
\begin{equation}
F=\sigma\frac{1+\varepsilon}{1+\alpha},
\label{eq:friction-final}
\end{equation}
but this is an external no-effective-voice envelope: $\Delta=\sigma$ requires $v_i=0$ for every party with positive stake. Under the equality constraint it is feasible only when $V_0=0$, when all required voice can be assigned to zero-stake parties, or when voice can be assigned outside the affected set. Otherwise the canonical index is not the program's optimum.

The conclusion is deliberately limited. The ratio is compatible with a stipulated coordination-cost story; it is not selected by the Lagrangian, uniquely implied by monotonicity, or validated by the existence of a constrained program. AOC owns the full index motivation and its underdetermination ledger at \texttt{TS-FORM}; the ROM technical supplement does not duplicate that layer.

If a future application needs a genuine allocation model, it should start from the actors, feasible set, and welfare or authorization criterion, then derive the implied response surface. The canonical ratio can be included as one prespecified candidate in that comparison. It should not be placed in the objective and then presented as though the optimization rediscovered it.

\subsection{Information-theoretic measurement motivation and its limit}
\label{app:information-motivation}

Information theory supplies candidate measurements for preference--decision dependence and information loss. It does not derive the canonical friction ratio, and it cannot recover the sign of alignment from mutual information alone.

Let $\mathbf U=(U_1,\ldots,U_n)$ denote the vector of affected-party preference sources and $A$ the controller's realized decision. For $H(\mathbf U)>0$, define normalized dependence strength
\begin{equation}
D_{UA}:=\frac{I(\mathbf U;A)}{H(\mathbf U)}\in[0,1].
\label{eq:normalized-preference-action-dependence}
\end{equation}
The statistic is sign-blind. A faithful invertible map and a systematic inversion can carry the same mutual information. Accordingly, the signed coordinate $\alpha\in(-1,1]$ used by the friction index is a separate directional construct: its sign must come from an ordered outcome, a declared preference--decision relation, or another measurement model. Setting $|\alpha|=D_{UA}$ is at most a magnitude gloss; it is not an identity supplied by information theory.

The Bayesian-update correspondence does not repair this gap. A normalized selection update and Bayes' rule share an algebraic form after fitness is assigned the role of likelihood. Mutual information, by contrast, is an expectation under a specified joint distribution and prediction experiment. A system can implement the normalized algebra without maximizing mutual information, and a high-information channel can transmit a decision rule opposed to the affected party's ordering. Direction, calibration, and informativeness are separate measurement axes.

Lagged predictive dependence can instead be summarized by transfer entropy,
\[
T_{\mathbf U\to A}=I(\mathbf U_t;A_{t+1}\mid A_t),
\]
but it remains nonnegative and does not identify a causal effect or alignment sign. It is useful only after the temporal ordering, conditioning history, estimator, and null model have been declared.

An information-loss coordinate also requires a convention. For example, under a declared partial-information-decomposition measure, let $R(U_1,\ldots,U_n;A)$ denote information about $A$ that is redundant across the component preference sources. One might define
\begin{equation}
\varepsilon:=1-\frac{R(U_1,\ldots,U_n;A)}{I(\mathbf U;A)}
\label{eq:epsilon-def}
\end{equation}
when $I(\mathbf U;A)>0$ and the selected redundancy measure obeys $0\leq R\leq I$. This quantity is undefined at $I=0$, depends on the redundancy definition, and does not by itself show that an institution lost information. Other applications may use calibrated observation error, posterior uncertainty, disclosure gaps, or misperception measures instead. These alternatives need not be numerically commensurable.

Even within one convention, normalization matters. Dividing by $H(\mathbf U)$ answers how much uncertainty about preference is removed by observing action; dividing by $H(A)$ answers a different question; an unnormalized $I(\mathbf U;A)$ changes with the entropy of the task. Comparisons across domains therefore require common support or an explicit adjustment. Estimation bias can also be large when the action alphabet is sparse, histories are dependent, or some preferences are unobserved. Those are properties of the data-generating process, not an abstract entropy force.

The canonical ratio can be read as a cost per stipulated unit of successful transmission only by assigning information demand proportional to $1+\varepsilon$ and effective capacity proportional to $1+\alpha$. That assignment gives
\[
\sigma\frac{\text{assigned demand}}{\text{assigned capacity}}
=\sigma\frac{1+\varepsilon}{1+\alpha},
\]
but the numerator, denominator, and reciprocal composition are modeling choices. Rate--distortion theory would require an explicit source distribution and distortion measure; mechanism design would require an objective and incentive constraints. Neither is supplied by the ratio alone.

This distinction matters empirically. The signed MARL control battery described in Section~\ref{sec:evidence-status} finds a positive mean observation-noise penalty in one feasible-centred shared-state environment, but the penalty grows toward more positive target-coordinate correlation rather than toward opposition. Thus the proxy mean direction can survive while the ratio's predicted alignment--noise interaction is contradicted in that environment. Information-theoretic vocabulary should therefore guide measurement design, not shield the functional form from discrimination.

An application should consequently record at least five choices: the source variable, decision variable, temporal ordering, conditioning set, and estimator. It should then validate the information coordinate against the actual perturbation. Observation noise, missing disclosure, strategic concealment, and preference heterogeneity can all reduce a statistic for different reasons. Treating them as one $\varepsilon$ is justified only if the intended prediction is stable across those interventions or the model explicitly separates them.

The full AOC measurement record retains the longer channel-capacity and proxy catalogue at \texttt{TS-MEASURE}. ROM keeps only this compact interface because information-theoretic motivation neither derives the index nor establishes its empirical fit.

\section{The ladder constraint}

\subsection{The Ladder Constraint: Exact Closure and Stepwise Agreement}
\label{app:ladder-constraint}

The defensible content of the Ladder Constraint is an exact-closure statement from standard Markov-chain lumpability theory. If each adjacent projection is strongly lumpable, direct and stepwise coarse descriptions agree exactly. If a projection is not lumpable, no single first-order Markov kernel reproduces the projected dynamics for every initial micro-distribution. These results do not order the approximation errors of direct and stepwise models outside the exact case.

\subsubsection{Definitions}

\begin{definition}[Nested scale hierarchy]
\label{def:scale-hierarchy}
Let $T_0,T_1,T_2$ be finite state spaces with surjections $q_{01}:T_0\to T_1$ and $q_{12}:T_1\to T_2$. Their composite is $q_{02}=q_{12}\circ q_{01}$.
\end{definition}

\begin{definition}[Pushforward]
\label{def:cg-operator}
For a distribution $p$ on $T_a$, the pushforward under $q_{ab}$ is
\[
(\Pi_{ab}p)(B):=\sum_{x:q_{ab}(x)=B}p(x).
\]
Pushforwards compose exactly: $\Pi_{02}=\Pi_{12}\Pi_{01}$.
\end{definition}

\begin{definition}[Transition dynamics]
\label{def:transition-dynamics}
Let $M_0$ be a row-stochastic kernel on $T_0$. In discrete time, row distributions evolve as $p_{t+1}=p_tM_0$. The statements below concern this transfer slice. A full \ROM{} quotient additionally requires the effective-fitness rate $\phi=w\rho$ to factor through the relevant projection.
\end{definition}

\begin{definition}[Strong lumpability and quotient kernel]
\label{def:lumpability}
The kernel $M_0$ is strongly lumpable with respect to $q_{01}$ if, for every pair $x,x'$ with $q_{01}(x)=q_{01}(x')$ and every block $B\in T_1$,
\[
\sum_{y:q_{01}(y)=B}M_0(x,y)
=
\sum_{y:q_{01}(y)=B}M_0(x',y).
\]
When this holds, the common value defines the quotient kernel
\begin{equation}
M_1(A,B):=\sum_{y:q_{01}(y)=B}M_0(x,y),
\qquad x\in q_{01}^{-1}(A).
\label{eq:surrogate}
\end{equation}
The definition is independent of the representative $x$ precisely because of strong lumpability.
\label{def:surrogate}
\end{definition}

\subsubsection{Exact stepwise agreement}

\begin{theorem}[Nested lumpability implies exact stepwise agreement]
\label{thm:ladder-constraint}
Suppose $M_0$ is strongly lumpable with respect to $q_{01}$, with quotient $M_1$, and $M_1$ is strongly lumpable with respect to $q_{12}$, with quotient $M_2$. Then $M_0$ is strongly lumpable with respect to the composite $q_{02}$. Its direct quotient is $M_2$, and for every initial distribution $p$ and every integer $t\geq0$,
\begin{equation}
\Pi_{02}(pM_0^t)
=\Pi_{12}\!\left((\Pi_{01}p)M_1^t\right)
=(\Pi_{02}p)M_2^t.
\label{eq:ladder-bound}
\end{equation}
Thus the exact direct and exact stepwise coarse descriptions agree; no intermediate-scale penalty or benefit arises in the lumpable case.
\end{theorem}

\begin{proof}
For $x\in T_0$ and $C\in T_2$, group the fine-state transition sum first by its $T_1$ blocks:
\[
\sum_{z:q_{02}(z)=C}M_0(x,z)
=\sum_{B:q_{12}(B)=C}M_1(q_{01}(x),B).
\]
If $q_{02}(x)=q_{02}(x')$, then $q_{01}(x)$ and $q_{01}(x')$ lie in the same $q_{12}$ block. Strong lumpability of $M_1$ therefore makes the displayed quantity identical for $x$ and $x'$, proving composite lumpability and identifying the direct quotient with $M_2$. The evolution identity follows by applying the defining intertwining relation $\Pi_{ab}(pM_a)=(\Pi_{ab}p)M_b$ repeatedly.
\end{proof}

The converse does not follow: a fine kernel may be lumpable under the composite partition even when a chosen intermediate partition is not lumpable. The theorem therefore characterizes one sufficient exact ladder, not a requirement that every valid coarse description expose all intermediate scales.

\subsubsection{What failure of lumpability establishes}

\begin{proposition}[Failure of universal first-order closure]
\label{prop:memory}
If $M_0$ is not strongly lumpable with respect to a projection $q$, there is no row-stochastic kernel $K$ on the coarse space such that
\[
\Pi_q(pM_0)=(\Pi_qp)K
\]
for every initial distribution $p$ on the fine space.
\end{proposition}

\begin{proof}
Failure of strong lumpability supplies two fine states $x,x'$ in the same coarse block and a destination block $B$ for which the aggregate transition probabilities differ. The point masses $\delta_x$ and $\delta_{x'}$ have the same coarse initial distribution, but their one-step coarse distributions assign different mass to $B$. A single coarse kernel $K$ cannot map the same coarse initial state to both outcomes.
\end{proof}

This proposition is a universal-closure result. For a restricted initial ensemble, a particular stationary distribution, or a finite observation window, a useful Markov approximation may still exist. An exact reduced representation can instead retain hidden state or history dependence; the Mori--Zwanzig formalism provides one such memory representation \citep{mori1965transport, zwanzig1961memory, aristoff2023coarsegraining}. Non-lumpability alone does not define a unique memory kernel or prove that every fitted coarse model has positive error.

\begin{remark}[Retraction of general error orderings]
\label{rem:retraction}
An earlier draft asserted a strict super-additive penalty for scale-skipping. A later draft replaced it with a general sub-additive ordering. Neither claim is retained. Outside exact lumpability, direct and stepwise approximations can use different state spaces, estimators, initial ensembles, horizons, and loss functions. Without fixing those choices there is no well-defined scalar comparison, much less a universal inequality. Equation~\eqref{eq:ladder-bound} is an equality for exact quotients, not an approximation-error bound.
\end{remark}

\subsubsection{Implication for institutional models}

For an institutional hierarchy, the theorem supplies a diagnostic rather than a constitutional prescription. If individual-to-community and community-to-national dynamics satisfy the required within-block transition and effective-fitness homogeneity, bypassing the intermediate representation changes nothing in the exact projected dynamics. If composite lumpability fails, a memoryless national kernel cannot represent every individual-level initial condition; an intermediate state, another latent variable, or explicit memory may improve fidelity. Whether it actually improves prediction must be measured for the chosen approximation. Intermediate institutions are therefore possible carriers of omitted state, not mathematically mandatory steps and not guaranteed error reducers.

\noindent\textit{Note on scope.} The exact theorem above is standard finite-state lumpability. Mean-field limits, time-scale separation, symmetry, and renormalization arguments may justify approximate closure, but each requires its own asymptotic assumptions and error analysis. Appendix~\ref{app:ladder-exceptions} treats them only as candidate regimes, not automatic exceptions.


\subsection{Candidate Regimes for Exact or Approximate Closure}
\label{app:ladder-exceptions}

Appendix~\ref{app:ladder-constraint} establishes one exact result: nested strong lumpability gives stepwise agreement, while failure of lumpability rules out a universal first-order coarse kernel. The regimes below can make exact or approximate closure plausible, but none supplies a general ordering between direct and stepwise approximation errors. Each application must check the actual transition and effective-fitness maps.

\begin{remark}[Renormalization-group fixed points]
\label{rem:rg-fixed}
At a renormalization-group fixed point, repeated coarse-graining preserves a functional form up to parameter rescaling. That property can support an exact scale map when the coarse-graining operator also intertwines the dynamics. Scale invariance alone, however, does not imply Markov lumpability or a vanishing memory term for an arbitrary observable. A \ROM{} application must specify the state map and verify closure rather than infer it from critical scaling.
\end{remark}

\begin{remark}[Mean-field limits]
\label{rem:mean-field}
When interactions are exchangeable and depend only on an aggregate statistic, that statistic may close exactly in an infinite-population limit or approximately for large finite populations. The relevant condition is not merely that the network is dense: transition probabilities and effective-fitness rates must depend on micro-states only through the retained aggregate. Propagation-of-chaos or concentration arguments are needed to quantify a finite-population approximation.
\end{remark}

\begin{remark}[Time-scale separation]
\label{rem:timescale}
If unresolved variables mix much faster than retained variables change, conditional equilibration can yield an approximately autonomous slow process. This is an approximation regime, not exact lumpability by definition. Its error depends on the ratio of time scales, spectral properties of the fast process, the observation horizon, and the initial departure from conditional equilibrium. A claim that an intermediate scale can be omitted therefore requires a singular-perturbation or averaging bound for the specified model.
\end{remark}

\begin{remark}[Criticality]
\label{rem:criticality}
Criticality is not a generic exception to the closure problem. Long-range correlations and slow modes can make local Markov closure less accurate even when observables exhibit scaling laws. A critical fixed point may provide a controlled renormalized description for selected observables, but it does not license bypassing arbitrary intermediate institutional levels. No emergency-governance conclusion follows from criticality alone.
\end{remark}

\begin{remark}[Symmetry and homogeneous blocks]
\label{rem:symmetry}
Exact symmetry can imply lumpability when the kernel is equivariant under permutations within each block and effective-fitness rates are constant on those blocks. Exchangeability of one observed distribution is insufficient: the transition law itself must respect the partition. Approximate homogeneity can motivate a reduced model, but the residual within-block variation must be included in its error analysis.
\end{remark}

\subsubsection*{Design implication}

Standardization, fungibility, and stable time-scale separation can make a coarse institutional model easier to defend because they reduce within-block heterogeneity. Conversely, local interactions and slow unresolved modes warn against a memoryless aggregate description. These are modeling diagnostics, not proofs that centralization, federalism, entrenchment, or emergency authority is legitimate. The appropriate procedure is to state the proposed partition, test transition and effective-fitness homogeneity, and compare candidate coarse models under a common loss, initial ensemble, and horizon.

\subsubsection*{The general rule}

The general rule is therefore conditional: exact scale-skipping is safe when the composite projection is lumpable for the dynamics of interest; approximate scale-skipping is useful when a specified reduced model meets a declared error tolerance. When neither has been shown, the status is unresolved. Coarse-graining may omit memory or hidden state, but there is no theorem that it necessarily accumulates more error or that inserting every intermediate scale necessarily improves prediction.

\section{Network topology: scope diagnostics}
\label{app:network-topology}

The interaction structure $G_{S,t}$ can affect effective fitness, transfer, and the adequacy of a coarse state. This appendix records diagnostics rather than a universal correction. Results from a particular graph, game, or update rule do not transfer to ROM merely because both use evolutionary notation.

Graph evolutionary models demonstrate the possible range. Under their respective assumptions, regular-graph rules, heterogeneous-degree effects, and amplifier or suppressor topologies can alter fixation and cooperation comparisons \citep{ohtsuki2006simple,santos2005scale,lieberman2005evolutionary}. For a ROM application these results motivate retaining network position when it predicts transition or effective-fitness differences within an otherwise common type. They do not supply a generic additive term $\delta\rho(G)$.

\begin{center}
\begin{tabular}{p{0.27\textwidth}p{0.31\textwidth}p{0.31\textwidth}}
\toprule
 & \textbf{A well-mixed comparator is plausible} & \textbf{Retain or test network state} \\
\midrule
Connectivity & Dense or rapidly mixing contacts & Sparse, clustered, modular contacts \\
Update rule & Aggregate or self-evaluation rule & Local comparison or edge-dependent payoffs \\
Heterogeneity & Weak residual degree effects & Degree or centrality predicts transitions \\
Dynamics & Fixed or exogenous topology & Endogenous rewiring or partner choice \\
Layers & One layer or weak coupling & Coupled economic, social, and information layers \\
\bottomrule
\end{tabular}
\end{center}

None of the left-hand conditions alone proves a well-mixed approximation. The proposed aggregate state must still be compared with a network-resolved model over a common initial ensemble, horizon, and loss. Conversely, a right-hand diagnostic does not prove that aggregation fails; it identifies omitted state worth testing.

There are three distinct places for topology to enter. It can change the rate at which a source type contributes mass, in which case it belongs in $\phi=w\rho$; it can change the destination conditional on contribution, in which case it belongs in $M$; or it can change the type coding itself, for example when a ``broker'' is defined by network position. These choices are not algebraically interchangeable. Collapsing all three into a scalar network adjustment obscures both identification and the closure test.

Pair or moment closures enlarge the state from type frequencies to local configurations. Their hierarchy creates its own closure problem: pair dynamics depend on higher-order neighborhoods. Adaptive networks require a coupled specification,
\begin{equation}
\dot p=f(p,G),\qquad \dot G=g(p,G),
\label{eq:coevolution-diagnostic}
\end{equation}
but Equation~\eqref{eq:coevolution-diagnostic} is not part of ROM unless an application supplies $g$. Writing $G_{S,t}$ as an input in Equation~\eqref{eq:rom-main} permits a time-varying environment; it does not endogenize the network.

The exact closure result remains strong lumpability. If unresolved network positions give two microstates in the same block different block-to-block transition sums, the chosen projection is not strongly lumpable. A universal first-order quotient then fails, although an enlarged state, a history-dependent representation, or an approximation for a restricted ensemble may still work. Multi-layer aggregation has the same burden: independent layer reduction is justified only when the joint transition and effective-fitness maps factor through the retained variables.

Network aggregation can also preserve one quantity while losing another. A quotient fitted to reproduce one-step block transitions need not reproduce fixation probabilities, path-dependent exposures, or response to a targeted intervention. Those tasks weight hidden positions differently. Model comparison should therefore use the outcome and horizon named in the main estimand rather than a generic reconstruction score.

A practical audit asks four questions:
\begin{enumerate}[noitemsep]
    \item Does mixing occur faster than the observation horizon?
    \item After conditioning on declared type, does degree, community, or layer membership predict transition or effective fitness?
    \item Are ties selected in response to strategies or outcomes?
    \item Does a network-resolved comparator improve held-out prediction enough to justify its additional state?
\end{enumerate}
The answer determines whether $G$ can be treated as fixed context, must enter $\rho$ or $M$, or should become an explicit dynamical state. It does not determine the sign of the omitted-network bias in advance.

If ties rewire in response to $p$, the pair $(p,G)$ is the minimal candidate autonomous state; projecting to $p$ alone is generally non-Markov. If topology changes exogenously, the system is nonautonomous and the static Perron theorem does not apply even when each instantaneous kernel is primitive. If topology is fixed, heterogeneity can still prevent type-level closure. These cases require different analyses and should not be grouped under the phrase ``network effects.''


\section{Potential Structure and Kernel Reversibility: Scoped Results}
\label{app:gradient-flow}

This appendix separates three questions that earlier drafts conflated: when a pure-selection replicator field is a gradient, whether the proposed ownership tilt preserves reversibility, and what a stationary probability current does and does not imply. No general free-energy, quasi-potential, convergence, or cycle theorem is asserted for the full state-dependent replicator--mutator equation.

\subsection{Pure-selection potential structure}

\begin{definition}[Shahshahani metric]
\label{def:shahshahani}
On the interior of the probability simplex $\Delta_n$, the Shahshahani metric is
\begin{equation}
g_{ij}(p)=\frac{\delta_{ij}}{p_i}.
\label{eq:shahshahani}
\end{equation}
\end{definition}

\begin{definition}[Potential fitness field]
\label{def:potential-game}
A continuously differentiable fitness field $f=(f_1,\ldots,f_n)$ is potential on the simplex interior if there is a continuously differentiable scalar $\mathcal P$ such that, up to a common component that cancels under the simplex projection,
\begin{equation}
\frac{\partial\mathcal P}{\partial p_i}=f_i(p).
\label{eq:potential-condition}
\end{equation}
A sufficient local integrability condition is the symmetry of cross-partials,
\begin{equation}
\frac{\partial f_i}{\partial p_j}=\frac{\partial f_j}{\partial p_i}.
\label{eq:symmetry-condition}
\end{equation}
\end{definition}

\begin{theorem}[Pure-selection potential slice]
\label{thm:hs-gradient}
If $f$ has potential $\mathcal P$, then the pure-selection replicator equation
\begin{equation}
\dot p_i=p_i\bigl(f_i(p)-\bar f(p)\bigr),
\qquad \bar f(p)=\sum_jp_jf_j(p),
\label{eq:replicator-standard}
\end{equation}
is the Shahshahani gradient ascent of $\mathcal P$:
\begin{equation}
\dot p=\nabla^{\mathrm{Shah}}\mathcal P(p).
\label{eq:gradient-flow}
\end{equation}
\end{theorem}

\begin{proof}
The tangent projection of the Shahshahani gradient has component
\[
p_i\left(\frac{\partial\mathcal P}{\partial p_i}
-\sum_jp_j\frac{\partial\mathcal P}{\partial p_j}\right).
\]
Substitution of Equation~\eqref{eq:potential-condition} gives Equation~\eqref{eq:replicator-standard}. This is the standard potential-game slice of replicator dynamics \citep{shahshahani1979new, hofbauer1998evolutionary, monderer1996potential}.
\end{proof}

For frozen, type-specific effective-fitness rates $\phi_i$ that do not depend on $p$, the linear potential
\begin{equation}
\mathcal P_{\mathrm{frozen}}(p)=\sum_i p_i\phi_i
\label{eq:rom-potential}
\end{equation}
generates the pure-selection dynamics. In the bounded consent instantiation, $\phi_i=w_iL_i/(1+F_i)$, so
\begin{equation}
\mathcal P_{\mathrm{consent}}(p)=\sum_i p_i\frac{w_iL_i}{1+F_i}.
\label{eq:consent-potential}
\end{equation}
This yields a narrow comparative static: holding $w_i$ and $L_i$ fixed, larger $F_i$ lowers type $i$'s effective-fitness rate. It does not establish a potential for frequency-dependent fitness, mutation, or a time-varying network, and it ranks friction alone only when the other factors are controlled.

\begin{corollary}[Frozen consent pure-selection slice]
\label{cor:consent-gradient}
With $M=I$ and time-independent $w_i,L_i,F_i$, the consent-weighted replicator equation is the Shahshahani gradient ascent of Equation~\eqref{eq:consent-potential}.
\end{corollary}

\subsection{Exact reversibility of the normalized ownership tilt}

The full \ROM{} equation is restated only to mark scope:
\begin{equation}
\dot p_i=\sum_jp_j\phi_j(p,t)M_{ji}-p_i\bar\phi.
\label{eq:rom-restate}
\end{equation}
The following theorem concerns the transfer kernel $M$ alone. It is not a gradient theorem for Equation~\eqref{eq:rom-restate}.

\begin{theorem}[Reversibility under a normalized destination tilt]
\label{thm:rom-gradient}
\label{thm:ownership-tilt}
Let $M^0$ be a finite row-stochastic kernel reversible with respect to a positive measure $\mu$:
\begin{equation}
\mu_iM^0_{ij}=\mu_jM^0_{ji}.
\label{eq:detailed-balance}
\end{equation}
For a scalar state score $O_i$, define
\[
Z_i:=\sum_kM^0_{ik}e^{\gamma O_k},
\qquad
M_{ij}:=\frac{M^0_{ij}e^{\gamma O_j}}{Z_i}.
\]
Then $M$ is row-stochastic and reversible with respect to
\[
\pi_i:=\frac{\mu_i e^{\gamma O_i}Z_i}
{\sum_\ell\mu_\ell e^{\gamma O_\ell}Z_\ell}.
\]
\end{theorem}

\begin{proof}
Row stochasticity follows from the definition of $Z_i$. Writing the normalizing denominator of $\pi$ as $C$,
\[
\pi_iM_{ij}
=\frac{\mu_i e^{\gamma O_i}Z_i}{C}
  \frac{M^0_{ij}e^{\gamma O_j}}{Z_i}
=\frac{\mu_iM^0_{ij}e^{\gamma(O_i+O_j)}}{C}
=\frac{\mu_jM^0_{ji}e^{\gamma(O_j+O_i)}}{C}
=\pi_jM_{ji}.
\]
\end{proof}

The original factor $\exp[-\gamma(O_i-O_j)]$ gives this same kernel after row normalization because the source factor $e^{-\gamma O_i}$ cancels. Consequently, unequal ownership scores do not make the proposed scalar tilt nonreversible when the baseline kernel is reversible. Nonreversibility would require a baseline current or a genuinely edge-dependent modulation that cannot be written as a scalar destination tilt.

\subsection{Scoped counterexamples}

\subsubsection{Frequency dependence can be non-potential}

For rock--paper--scissors, let
\begin{equation}
A=\begin{pmatrix}0&-1&1\\1&0&-1\\-1&1&0\end{pmatrix}.
\label{eq:rps-matrix}
\end{equation}
With $f(p)=Ap$, the cross-partials are not symmetric. The pure-selection replicator field is therefore not generated by the potential of Theorem~\ref{thm:hs-gradient}; in the zero-sum interior case it has closed level sets rather than monotonic ascent to a strict maximizer \citep{hofbauer1998evolutionary}. This example shows that a specified cyclic payoff can generate cycles. It does not show that generic network dependence, or the consent-friction specification in particular, does so.

\subsubsection{Stationary current is not a population limit cycle}

Consider the row-stochastic kernel
\begin{equation}
\begin{aligned}
&M_{12}=M_{23}=M_{31}=0.3,\\
&M_{21}=M_{32}=M_{13}=0.1,\\
&M_{11}=M_{22}=M_{33}=0.6.
\end{aligned}
\label{eq:asymmetric-mutation}
\end{equation}
The uniform distribution is stationary, but Kolmogorov's cycle condition fails:
\begin{equation}
M_{12}M_{23}M_{31}=0.027\neq0.001=M_{13}M_{32}M_{21}.
\label{eq:kolmogorov-violation}
\end{equation}
At stationarity the edge current is
\begin{equation}
J_{i\to i+1}=\tfrac13(0.3)-\tfrac13(0.1)=\tfrac1{15}>0.
\label{eq:stationary-current}
\end{equation}
This kernel is irreducible and aperiodic, so its mass distribution converges to the uniform stationary distribution even though the stationary edge current is nonzero. The example distinguishes circulation of transition flow from oscillation of $p_t$: nonreversibility alone does not imply a limit cycle, an attractor orbit, or chaos.

\subsection{Implications and scope}

The established claims are therefore limited. Pure selection is a Shahshahani gradient when its fitness field has a potential. The normalized ownership tilt preserves reversibility whenever its baseline kernel is reversible. Other kernels or frequency-dependent fitness fields may be nonreversible or non-potential, but their long-run behavior must be analyzed from the specified autonomous system. If $G_t$, $\phi(p,t)$, or $M_t$ changes with time, a time-independent potential or stationary-distribution argument is inapplicable without an additional nonautonomous theorem. The present paper supplies no general convergence, tracking, cycling, or chaos result for that setting.

\section{Ownership-sensitive transfer: interpretation and test}
\label{app:ownership-microfoundations}

For a baseline row-stochastic kernel $M^0$ and scalar destination score $O_j$, the proposed normalized tilt is
\begin{equation}
M_{ij}=\frac{M^0_{ij}e^{\gamma O_j}}{Z_i},
\qquad Z_i=\sum_kM^0_{ik}e^{\gamma O_k},\qquad \gamma\geq0.
\label{eq:ownership-normalized}
\end{equation}
The source factor in the original expression $e^{-\gamma(O_i-O_j)}$ cancels under row normalization. The implemented object is therefore a destination tilt, not an unnormalized tenure law.

The normalization yields two directly testable identities. For two destinations $j$ and $k$ with positive baseline entries,
\begin{equation}
\log\frac{M_{ij}}{M_{ik}}
=\log\frac{M^0_{ij}}{M^0_{ik}}+\gamma(O_j-O_k),
\label{eq:ownership-log-odds}
\end{equation}
so the score difference has a source-invariant log-odds slope after conditioning on the baseline. The full sensitivity is
\begin{equation}
\frac{\partial M_{ij}}{\partial O_\ell}
=\gamma M_{ij}\bigl(\mathbf1\{j=\ell\}-M_{i\ell}\bigr).
\label{eq:ownership-softmax-derivative}
\end{equation}
Raising one destination score increases its probability and decreases the others within the same source row. It does not multiply every transition by an unnormalized persistence factor.

Several literatures can motivate this family only conditionally. A Boltzmann reading requires $O_j$ to enter a specified destination energy; a barrier-crossing reading requires it to lower a stated transition barrier; the endowment effect, loss aversion, and status-quo bias motivate an ownership-sensitive state variable but do not select an exponential without a response rule; and quantal response requires additive ownership utility and a baseline absorbing the other utilities \citep{castellano2009statistical,hanggi1990reaction,kahneman1979prospect,strahilevitz1998effect,samuelson1988status,mckelvey1995quantal}. Agreement on an exponential form is not convergent validation because the variables and sampling processes differ.

Appendix~\ref{app:gradient-flow} proves the exact kernel result: if $M^0$ is reversible with respect to a positive measure $\mu$, Equation~\eqref{eq:ownership-normalized} is reversible with respect to $\pi_i\propto\mu_i e^{\gamma O_i}Z_i$. The scalar tilt therefore preserves, rather than generically breaks, reversibility of a reversible baseline.

That theorem is one-way with respect to the baseline assumption. If $M^0$ already carries stationary current, the scalar tilt need not remove it. If the modulation is edge-specific rather than the gradient of a scalar destination score, the detailed-balance proof no longer goes through. Reversibility must then be tested from cycle products or pairwise stationary flows. In either case, kernel reversibility concerns transfer conditional on a source; it is not a convergence theorem for a state-dependent weighted ROM population.

A narrow comparative static survives. If $M^0_{ii}>0$, the baseline and all alternative scores are fixed, and $q_i:=1-M_{ii}$ is the one-step exit probability, then
\begin{equation}
\frac{\partial q_i}{\partial O_i}
=-\gamma M_{ii}(1-M_{ii})\leq0.
\label{eq:tenure-prediction}
\end{equation}
Strict inequality requires $\gamma>0$ and positive probability of an alternative destination. The second derivative changes sign at $M_{ii}=1/2$, so there is no global convex-resistance result. Calling $q_i$ a tenure hazard additionally requires a measured map from tenure to $O_i$, a stable baseline, and a link from the discrete kernel to observed events.

An empirical application should estimate $M^0$, define $O$, and compare the exponential tilt out of sample with untilted, power-law, saturating, and flexible destination-score models. Multidimensional, actor-specific, or edge-specific ownership can change both reversibility and the comparative static.

The cleanest test uses repeated source--destination transitions with exposure times and a score recorded before the transition. Equation~\eqref{eq:ownership-log-odds} can then be compared with source-specific slopes, nonlinear score effects, and an edge model. If $O$ is itself inferred from tenure after the event, the negative exit association is partly definitional and cannot identify the proposed mechanism. The baseline kernel should also be estimated from a period or control in which the ownership tilt is absent, or jointly identified under restrictions; calling the residual ``baseline'' does not make it exogenous.

\section{Formal verification in Lean 4}
\label{sec:appendix-lean}

The algebraic core has been checked in Lean~4 v4.27.0 with Mathlib v4.27.0. Twenty-two machine-checked theorems correspond directly to results named in this paper. Six further theorems from the companion \texttt{ROMEthics} welfare bridge are listed under a separate heading for completeness: this paper names no welfare or ethical-survival result, and one of the six was withdrawn as evidence on 30 July 2026 (see below). With eight further unlisted results, the current six modules contain 36 theorems in total, with zero \texttt{sorry}, zero \texttt{admit}, zero custom axioms, and no \texttt{decide}/\texttt{native\_decide} shortcuts. The table is the claim boundary: theorem names should not be read as proving more than their descriptions.

Three levels of assurance must be separated. Lean checks that the encoded theorem follows from its encoded hypotheses. The paper-to-code map checks that the encoding matches the displayed claim. Neither step shows that an empirical variable realizes the encoded scalar or that the assumptions hold in an application. The theorem count is therefore an inventory, not a strength metric; several results are deliberate boundary or definitional lemmas.

\newcommand{\ltt}[1]{{\ttfamily\footnotesize\def\_{\textunderscore\allowbreak}#1}}
\begin{longtable}{p{5.6cm}p{7.2cm}}
\caption{Machine-checked results mapped to the reader paper, followed by the unmapped companion welfare bridge.}\label{tab:lean-theorems-rom}\\
\toprule
\textbf{Lean theorem} & \textbf{Checked content} \\
\midrule
\endfirsthead
\toprule
\textbf{Lean theorem} & \textbf{Checked content} \\
\midrule
\endhead
\multicolumn{2}{l}{\textit{ROM/Basic.lean}}\\
\ltt{row\_stochastic\_sum} & Row-stochastic transfer preserves the total source mass.\\
\ltt{rom\_simplex\_invariant} & The derivative has zero total mass; despite the historical name, this is not nonnegative simplex invariance.\\
\ltt{identity\_kernel\_selection} & The identity-kernel algebra reduces to pure selection; no Bayesian or convergence theorem.\\
\ltt{identity\_kernel\_row\_stochastic} & The identity kernel is row-stochastic.\\
\ltt{detailed\_balance\_zero\_flow} & Pairwise detailed balance implies zero net pair flow.\\
\ltt{asymmetric\_kernel\_no\_uniform\_balance} & Unequal kernel entries remain unequal under a common positive factor.\\
\ltt{rps\_skew\_symmetric} & The displayed rock--paper--scissors matrix is skew-symmetric.\\
\ltt{rps\_not\_potential} & The displayed RPS field violates the integrability condition.\\
\ltt{skew\_symmetric\_nonzero\_not\_potential} & A nonzero skew entry rules out the stated potential symmetry.\\
\ltt{consent\_survival\_mono\_legitimacy} & The implemented scalar score $L/(1+F)$ rises with its first argument.\\
\ltt{consent\_survival\_anti\_friction} & The implemented scalar score falls with $F$.\\
\midrule
\multicolumn{2}{l}{\textit{ROM/Advanced.lean}}\\
\ltt{consent\_survival\_nonneg} & Nonnegativity under $L\geq0,F\geq0$.\\
\ltt{consent\_survival\_zero\_legitimacy} & $L=0$ gives zero score.\\
\ltt{consent\_survival\_scale\_legitimacy} & Linearity in the first scalar argument.\\
\ltt{consent\_survival\_le\_legitimacy} & $L/(1+F)\leq L$ for $F\geq0$.\\
\ltt{consent\_survival\_pos} & Positivity for $L>0,F\geq0$.\\
\ltt{row\_stochastic\_sum\_const} & Constant source propagated through a row-stochastic kernel.\\
\midrule
\multicolumn{2}{l}{\textit{ROM/Transfers.lean}}\\
\ltt{rom\_movingEquilibrium} & A defined shift-update emits the next point of its supplied path.\\
\ltt{rom\_no\_static\_if\_path\_varies} & A nonconstant sequence is not a constant fixed point.\\
\ltt{rom\_path\_boundedChase\_zero} & A path compared with itself has zero discrepancy.\\
\ltt{rom\_dissensus\_of\_positive\_discrepancy} & Positive stipulated discrepancy implies the defined dissensus inequality.\\
\ltt{romPath\_nonneg} & Nonnegative scalar inputs give a nonnegative defined path.\\
\midrule
\multicolumn{2}{l}{\textit{ROMEthics/} --- checked in the formalization, not claimed in this paper}\\
\ltt{ethicalSurvival\_mono\_benefit} & The implemented ethical score rises with benefit.\\
\ltt{ethicalSurvival\_anti\_harm} & The implemented ethical score falls with harm.\\
\ltt{ethicalSurvival\_eq\_rom\_survival\_of\_zero\_harm} & Zero harm recovers the implemented ROM score.\\
\ltt{ethicalSurvival\_mono\_alignment\_via\_friction} & Algebra only: a larger $\alpha$ lowers the stipulated $F=\sigma(1+\varepsilon)/(1+\alpha)$ and so raises the implemented score. \emph{Withdrawn 2026-07-30 as evidence}: the friction functional it depends on was retracted, and the theorem is not evidence that better alignment raises survival.\\
\ltt{ethicalSurvival\_eq\_welfare\_of\_zero\_friction} & Zero friction recovers the defined welfare score.\\
\ltt{ethicalSurvival\_le\_welfare} & The implemented ethical score does not exceed the defined welfare score.\\
\bottomrule
\end{longtable}

The mass result deserves special caution. The source theorem historically named \ltt{rom\_simplex\_invariant} proves that the derivative coordinates sum to zero under the encoded row-sum hypothesis. It does not encode an ODE solution, a time interval, or the boundary argument needed for componentwise nonnegativity. The reader text therefore calls it a zero-total-mass identity even while preserving the source name for reproducibility.

The consent-score theorems operate on real scalars satisfying explicit inequalities. They certify monotonicity, nonnegativity, scaling, and upper bounds of the function $L/(1+F)$. They do not certify that $L$ measures legitimacy, that $F$ measures social friction, or that either is measured without error. The formal names containing \texttt{legitimacy} are legacy identifiers; the reader paper interprets their first argument as descriptive effective voice.

The Transfers results are structural or definitional, not asymptotic. The named moving-equilibrium result applies to an update constructed to shift along a supplied sequence. It does not prove contraction toward an independently defined moving target. Similarly, the bounded-chase and dissensus lemmas unfold stipulated discrepancy definitions; they do not establish that an observed system has bounded tracking error or persistent disagreement.

The six \texttt{ROMEthics} rows inventory the companion welfare-bridge algebra and support no claim in this paper, which names no welfare or ethical-survival result and does not certify the ethical bridge. One of them, \ltt{ethicalSurvival\_mono\_alignment\_via\_friction}, is discharged through a friction functional that was retracted on 30 July 2026; the source marks the declaration withdrawn, and its statement holds as algebra about $\sigma(1+\varepsilon)/(1+\alpha)$ and is not evidence that better alignment raises ethical survival. The declaration is retained in the ancillary module with its withdrawal notice so that the checked algebra stays on record.

The following load-bearing claims remain outside the machine-checked layer:
\begin{itemize}[noitemsep]
    \item differential-equation existence and nonnegative simplex invariance;
    \item the Perron--Frobenius convergence and uniform-residual ranking theorem in Section~\ref{sec:static-theorem};
    \item strong lumpability, nested quotient agreement, and all approximation claims;
    \item the Bayesian scope statement and ownership-tilt reversibility proof;
    \item identification of $w$, $\rho$, $M$, $L$, $F$, $\alpha$, or $\varepsilon$ from data; and
    \item the AOC normative bridge from affected-party authorization to legitimacy.
\end{itemize}
These are proved on paper where a proof is supplied, treated as empirical questions where data are required, or explicitly left open. The technical supplement contains selected excerpts; exact copies of all six modules, taken from the cited repository at commit \texttt{8e11333} (25 August 2026, the commit that records the withdrawal), are ancillary under \texttt{anc/lean/} with their SHA-256 hashes in \texttt{SOURCE\_MANIFEST.md}, while the repository retains pinned build instructions \citep{farzulla2026lean}. Removing listings from the reader PDF changes presentation, not verification status.

\bibliographystyle{plainnat}
\bibliography{references}

\end{document}